\documentclass[prd, showpacs,preprintnumbers,amsmath,amssymb,nofootinbib,aps]{revtex4}

\usepackage{bm}
\usepackage{amsfonts}
\usepackage{mathrsfs}
\usepackage{graphicx}
\usepackage{amsmath}
\usepackage{float}
\usepackage{braket}

\begin{document}

\newcommand{\sgn}{\operatorname{sgn}}
\newcommand{\hhat}[1]{\hat {\hat{#1}}}
\newcommand{\pslash}[1]{#1\llap{\sl/}}
\newcommand{\kslash}[1]{\rlap{\sl/}#1}
\newcommand{\lab}[1]{}
\newcommand{\iref}[2]{}
\newcommand{\sto}[1]{\begin{center} \textit{#1} \end{center}}
\newcommand{\rf}[1]{{\color{blue}[\textit{#1}]}}
\newcommand{\eml}[1]{#1}
\newcommand{\el}[1]{\label{#1}}
\newcommand{\er}[1]{Eq.\eqref{#1}}
\newcommand{\df}[1]{\textbf{#1}}
\newcommand{\mdf}[1]{\pmb{#1}}
\newcommand{\ft}[1]{\footnote{#1}}
\newcommand{\n}[1]{$#1$}
\newcommand{\fals}[1]{$^\times$ #1}
\newcommand{\new}{{\color{red}$^{NEW}$ }}
\newcommand{\ci}[1]{}
\newcommand{\de}[1]{{\color{green}\underline{#1}}}
\newcommand{\ke}{\rangle}
\newcommand{\br}{\langle}
\newcommand{\lb}{\left(}
\newcommand{\rb}{\right)}
\newcommand{\lbk}{\left[}
\newcommand{\rbk}{\right]}
\newcommand{\blb}{\Big(}
\newcommand{\brb}{\Big)}
\newcommand{\nn}{\nonumber \\}
\newcommand{\p}{\partial}
\newcommand{\pd}[1]{\frac {\partial} {\partial #1}}
\newcommand{\cd}{\nabla}
\newcommand{\cc}{$>$}
\newcommand{\bqa}{\begin{eqnarray}}
\newcommand{\eqa}{\end{eqnarray}}
\newcommand{\bqe}{\begin{equation}}
\newcommand{\eqe}{\end{equation}}
\newcommand{\bay}[1]{\left(\begin{array}{#1}}
\newcommand{\eay}{\end{array}\right)}
\newcommand{\eg}{\textit{e.g.} }
\newcommand{\ie}{\textit{i.e.}, }
\newcommand{\iv}[1]{{#1}^{-1}}
\newcommand{\st}[1]{|#1\ke}
\newcommand{\at}[1]{{\Big|}_{#1}}
\newcommand{\zt}[1]{\texttt{#1}}
\newcommand{\non}{\nonumber}
\newcommand{\m}{\mu}
\def\xa{{m}}
\def\xA{{m}}
\def\xb{{\beta}}
\def\xB{{\Beta}}
\def\xd{{\delta}}
\def\xD{{\Delta}}
\def\xe{{\epsilon}}
\def\xE{{\Epsilon}}
\def\xve{{\varepsilon}}
\def\xg{{\gamma}}
\def\xG{{\Gamma}}
\def\xk{{\kappa}}
\def\xK{{\Kappa}}
\def\xl{{\lambda}}
\def\xL{{\Lambda}}
\def\xo{{\omega}}
\def\xO{{\Omega}}
\def\xvp{{\varphi}}
\def\xs{{\sigma}}
\def\xS{{\Sigma}}
\def\xt{{\theta}}
\def\xvt{{\vartheta}}
\def\xT{{\Theta}}
\def \Tr {{\rm Tr}}
\def\CA{{\cal A}}
\def\CC{{\cal C}}
\def\CD{{\cal D}}
\def\CE{{\cal E}}
\def\CF{{\cal F}}
\def\CH{{\cal H}}
\def\CJ{{\cal J}}
\def\CK{{\cal K}}
\def\CL{{\cal L}}
\def\CM{{\cal M}}
\def\CN{{\cal N}}
\def\CO{{\cal O}}
\def\CP{{\cal P}}
\def\CQ{{\cal Q}}
\def\CR{{\cal R}}
\def\CS{{\cal S}}
\def\CT{{\cal T}}
\def\CV{{\cal V}}
\def\CW{{\cal W}}
\def\CY{{\cal Y}}
\def\BC{\mathbb{C}}
\def\BR{\mathbb{R}}
\def\BZ{\mathbb{Z}}
\def\sA{\mathscr{A}}
\def\sB{\mathscr{B}}
\def\sF{\mathscr{F}}
\def\sG{\mathscr{G}}
\def\sH{\mathscr{H}}
\def\sJ{\mathscr{J}}
\def\sL{\mathscr{L}}
\def\sM{\mathscr{M}}
\def\sN{\mathscr{N}}
\def\sO{\mathscr{O}}
\def\sP{\mathscr{P}}
\def\sR{\mathscr{R}}
\def\sQ{\mathscr{Q}}
\def\sS{\mathscr{S}}
\def\sX{\mathscr{X}}

\def\slz{SL(2,Z)}
\def\slr{$SL(2,R)\times SL(2,R)$ }
\def\ads{${AdS}_5\times {S}^5$ }
\def\adst{${AdS}_3$ }
\def\sun{SU(N)}
\def\ad#1#2{{\frac \delta {\delta\sigma^{#1}} (#2)}}
\def\bqf{\bar Q_{\bar f}}
\def\nf{N_f}
\def\sunf{SU(N_f)}

\def\dcirc{{^\circ_\circ}}

\author{Morgan H. Lynch}
\email{mhlynch@uwm.edu}
\affiliation{Leonard E. Parker Center for Gravitation, Cosmology and Astrophysics, Department of Physics, University of Wisconsin-Milwaukee,
P.O.Box 413, Milwaukee, Wisconsin 53201, USA}

\title{Accelerated quantum dynamics}
\date{\today}

\begin{abstract}
In this paper we establish a formalism for the computation of observables due to acceleration-induced particle physics processes. General expressions for the transition rate, multiplicity, power, spectra, and displacement law of particles undergoing time-dependent acceleration and transitioning into a final state of arbitrary particle number are obtained. The transition rate, power, and spectra are characterized by unique polynomials of multiplicity and thermal distributions of both bosonic and fermionic statistics. The acceleration-dependent multiplicities are computed in terms of the branching fractions of the associated inertial processes. The displacement law of the spectra predicts that the energy of the emitted particles is directly proportional to the accelerated temperature.
\end{abstract}


\maketitle

\section{Introduction}
As a first order approximation to quantum gravity, quantum field theory in curved spacetime predicts the creation of particles from the vacuum. Of particular importance is the creation of thermalized particles via the Parker [1], Hawking [2], and Unruh [3] effects. These particle production mechanisms have given considerable insight into the quantum properties of spacetimes which contain horizons. The Rindler horizon present in accelerated reference frames enables one to use the Unruh effect, and its associated thermal bath, to induce particle transitions, i.e. decays and excitations. Our current understanding of these processes begins with the original computation of Mueller [4] which showed that one could use acceleration to change the decay rate of weakly interacting particles as well as excite stable particles into more massive states. Matsas and Vanzella [5-7] extended the analysis to fermions, analytically computed the power emitted during the transition, and numerically computed the spectra of the final state particles. In a previous work [8], we used scalar fields to extend the formalism to encompass transitions with arbitrary final state multiplicity. In this paper, the analytic computation of both the power and the spectra for arbitrary multiplicity is carried out. We also import certain relevant inertial quantities, such as the branching fractions of various decays, into the formalism and compute the acceleration scale to select the relevant decay pathway, i.e. multiplicity. Moreover, with the Planckian spectra obtained, we compute the peak energy of the emitted particles via a generalization of Wien's displacement law. This establishes that the most probable energy of the emitted particles is peaked about the accelerated temperature. These results are punctuated by the entire analysis being carried out using a newly developed time-dependent formalism which agrees with the previous developments of Obadia and Milgrom [9], Kothawala and Padmanabhan [10], and Barbado and Visser [11]. The time dependence and ability to compute a wide class of observables developed in this paper establish a basic foundation for an acceleration-induced particle physics phenomenology with applications to highly accelerated systems e.g. the inflationary epoch of the early universe or perhaps the evaporation of black holes.

In this paper, Sec. II computes the generalized response function. Using a fully relativistic treatment, i.e. quantized fields, the response function is derived in a similar manner as one would in the case of constant acceleration. However, to extend the analysis to encompass time dependence, the generalized response function is expressed as the Fourier transform with respect to the rapidity rather than the proper time. This prescription allows for the use of the same mathematical treatment in computing observables as in the case of constant proper acceleration. Section III focuses on the time-dependent formalism utilized to compute various spacetime quantities for use in characterizing the accelerated motion, e.g. trajectories, spacetime intervals, and Lorentz gamma. These quantities are used to evaluate the Wightman functions and its variants which correspond to the relevant observables. In Sec. IV, the appropriate variants of the Wightman functions corresponding to the transition rate, power, and spectra are computed for arbitrary trajectories. Each is then evaluated along the appropriate time-dependent trajectory derived in the previous section. In Sec. V we evaluate the acceleration-induced transition rate. In doing so, we also derive the transition rate polynomials of multiplicity, characterized by an integer index, along with the associated bosonic thermal distribution. Section VI utilizes the transition rates to compute the acceleration scale that selects the dominant multiplicity based on the branching fractions of the relevant processes in the inertial limit. Specific limiting cases are considered which yield analytic results. The radiated power is computed in Sec. VII using the appropriate variant of the Wightman function. In doing so, we also obtain the power polynomials of multiplicity, characterized by a half-integer index, along with the associated fermionic thermal distribution. Section VIII contains the derivation of the energy spectra. The resulting Planck-like spectra are expressed in terms of the spectra polynomials of multiplicity, characterized by an integer index, and the associated bosonic thermal distribution at finite chemical potential. In Sec. IX the displacement law of the peak energy emitted is computed for arbitrary acceleration, multiplicity, and transition energy. The result generalizes Wien's displacement law, in that the peak energy is proportional to the accelerated temperature weighted by a numerically determined parameter. Final comments and a summary of the conclusions are presented in Sec. X. All calculations are performed using the natural units $\hbar = c = k_{B} = 1$.

\section{Generalized response function}
In this section we set up a formalism capable of computing a wide class of particle transitions regardless of the number of final state products [4,8]. To facilitate the analysis, we consider all particles to be scalars. Consider an initial Rindler particle moving along an arbitrary time-dependent accelerated trajectory and decaying into a final state containing $n_{R}$ Rindler particles and $n$ Minkowksi particles. This process is schematically written as

\bqa
\Psi_{i} \rightarrow_{a} \Psi_{1} + \Psi_{2} + \cdots + \Psi_{n_{R}} + \phi_{1} + \phi_{2} + \cdots + \phi_{n}.
\eqa

The initial and final Rindler particles are denoted by $\Psi_{j}$ and are used to describe any particle under acceleration. We consider the initial Rindler particle to be massive while the final state Rindler particles can be massless or massive in any combination. The massless Minkowski particles in the final state are denoted by $\phi_{k}$ and are used to describe any particles propagating along inertial trajectories. To describe these transitions, we consider their coupling to be described by the following general interaction action,

\bqe
\hat{S}_{I} = \frac{G_{n}}{ \sqrt{\kappa}}\int d^{4}x\sqrt{-g}\hat{\Psi}_{i}\prod_{r = 1}^{n_{R}}\hat{\Psi}_{r} \prod_{m = 1}^{n}\hat{\phi}_{m}.
\eqe

The coupling constant $G_{n}$ is labeled by the multiplicity and can be determined by fixing it to a known process in the inertial limit. We note that the coupling may be dimensionful depending on the transition in question. The $\frac{1}{ \sqrt{\kappa}}$ term is included to absorb the overall normalization of our Rindler states. The domain of integration is confined to the right Rindler wedge where the modes of our accelerated fields are defined. We denote the Fock states of our Rindler particles $\ket{\Psi_{j}}$ with the index $j$ characterizing their energies. As usual, we label our Minkowski states $\ket{\mathbf{k}_{\ell}}$ by their momenta. We use the notation $\ket{\prod_{i}^{n}\mathbf{k}_{i}} = \ket{\mathbf{k}_{1},\mathbf{k}_{2},\cdots \mathbf{k}_{n}}$ for the Fock states of both the Rindler and Minkowski fields. Note that we leave off subscripts denoting the Rindler and Minkowski Fock states and let the field operators imply the label, i.e. $\ket{\Psi_{j}}_{R} \rightarrow \ket{\Psi_{j}}$ and $\ket{\mathbf{k}_{i}}_{M} \rightarrow \ket{\mathbf{k}_{i}}$. Working in the interaction picture, the acceleration-induced probability amplitude for our massive initial state to transition into $N =n_{R}+n$ final state particles is given by,

\bqe
\mathcal{A} = \bra{\prod_{\ell = 1}^{n}\mathbf{k}_{\ell}}\otimes \bra{\prod_{j = 1}^{n_{R}}\Psi_{j}} \hat{S}_{I}\ket{\Psi_{i}} \otimes \ket{0}.
\eqe

The magnitude squared of the transition amplitude gives the differential probability per unit momenta of each Minkowski particle, i.e. $\frac{d \mathcal{P}}{D^{3}_{n}k} =  |\mathcal{A}|^{2}$. Note that we are using the more compact notation  $\prod_{j = 1}^{n} d^{3}k_{j} = D^{3}_{n}k$. Moreover, since we are taking the magnitude squared of a complex integral, we remind the reader that there are two dummy variables in expressions such as $|\int f(x)dx|^2 = \iint dxdx'|f(x)|^2$, where $|f(x)|^2 \equiv f(x)f^{\ast}(x')$. Thus the differential probability of our $N$-particle acceleration-induced transition is given by

\bqa
\frac{d \mathcal{P}}{D^{3}_{n}k} &=&  \frac{G_{n}^{2}}{ \kappa}\left| \int d^{4}x\sqrt{-g}\bra{\prod_{\ell = 1}^{n}\mathbf{k}_{\ell}}\otimes \bra{\prod_{j = 1}^{n_{R}}\Psi_{j}}   \hat{\Psi}_{i}(x) \prod_{r = 1}^{n_{R}}\hat{\Psi}_{r}(x) \prod_{m = 1}^{n}\hat{\phi}_{m}(x) \ket{\Psi_{i}} \otimes \ket{0} \right|^{2} \nonumber \\
&=& \frac{G_{n}^{2}}{ \kappa}\left| \int d^{4}x\sqrt{-g} \bra{\prod_{j = 1}^{n_{R}}\Psi_{j}}   \hat{\Psi}_{i}(x) \prod_{r = 1}^{n_{R}}\hat{\Psi}_{r}(x) \ket{\Psi_{i}} \bra{\prod_{\ell = 1}^{n}\mathbf{k}_{\ell}} \prod_{m = 1}^{n}\hat{\phi}_{m}(x)   \ket{0} \right|^{2} \non \\
&=& \frac{G_{n}^{2}}{ \kappa} \iint d^{4}x d^{4}x' \sqrt{-g}\sqrt{-g'} \left|\bra{\prod_{j = 1}^{n_{R}}\Psi_{j}}   \hat{\Psi}_{i}(x) \prod_{r = 1}^{n_{R}}\hat{\Psi}_{r}(x) \ket{\Psi_{i}}\right|^{2} \left| \bra{\prod_{\ell = 1}^{n}\mathbf{k}_{\ell}} \prod_{m = 1}^{n}\hat{\phi}_{m}(x)   \ket{0} \right|^{2}.
\eqa  

By factoring out the $n$ complete sets of momentum eigenstates we can further simplify the above expression. The resultant two-point functions, i.e. Wightman functions, characterize the probability for each Minkowski particle to propagate along the accelerated trajectory. The product of the $n$ Wightman functions then characterizes the total probability that all of the $n$ Minkowski particles simultaneously propagate together. Moreover, we will endow each Wightman function with the index $m$ to label each of the Minkowski particles and also the relevant observables computed. Hence

\bqa
\mathcal{P} &=&   \frac{G_{n}^{2}}{ \kappa} \iint d^{4}x d^{4}x' \sqrt{-g}\sqrt{-g'} \left|\bra{\prod_{j = 1}^{n_{R}}\Psi_{j}}   \hat{\Psi}_{i}(x) \prod_{r = 1}^{n_{R}}\hat{\Psi}_{r}(x) \ket{\Psi_{i}}\right|^{2} \prod_{n = 1}^{n} \int d^{3}k_{n} \left| \bra{\prod_{\ell = 1}^{n}\mathbf{k}_{\ell}} \prod_{m = 1}^{n}\hat{\phi}_{m}(x)   \ket{0} \right|^{2} \non \\
&=&  \frac{G_{n}^{2}}{ \kappa} \iint d^{4}x d^{4}x' \sqrt{-g}\sqrt{-g'} \left|\bra{\prod_{j = 1}^{n_{R}}\Psi_{j}}   \hat{\Psi}_{i}(x) \prod_{r = 1}^{n_{R}}\hat{\Psi}_{r}(x) \ket{\Psi_{i}}\right|^{2}  \prod_{m = 1}^{n} \bra{0}\hat{\phi}_{m}(x') \hat{\phi}_{m}(x)   \ket{0}\non \\
&=& \frac{G_{n}^{2}}{ \kappa} \iint d^{4}x d^{4}x' \sqrt{-g}\sqrt{-g'} \left|\bra{\prod_{j = 1}^{n_{R}}\Psi_{j}}   \hat{\Psi}_{i}(x) \prod_{r = 1}^{n_{R}}\hat{\Psi}_{r}(x) \ket{\Psi_{i}}\right|^{2}  \prod_{m = 1}^{n}G_{m}^{\pm}[x',x].
\eqa

The inner products over our Rindler fields in the above expression select the appropriate mode function of each particle. For Rindler particles [12], these mode functions can be written as $f_{\Psi}[x(\tau)]e^{-i\omega \tau}$. Analyzing the transition process in a frame comoving with the initial accelerated particle allows us to parametrize the system with its proper time. This implies that its energy is just its rest mass, i.e. $\omega_{i} = m_{i}$, while the final particles have an arbitrary energy $\omega_{r}$. The inner products then imply

\bqa
\bra{\prod_{j = 1}^{n_{R}}\Psi_{j}}   \hat{\Psi}_{i}(x) \prod_{r = 1}^{n_{R}}\hat{\Psi}_{r}(x) \ket{\Psi_{i}} &=& f_{\Psi_{i}}[x(\tau)]e^{-im_{i}\tau} \prod_{r = 1}^{n_{R}} f_{\Psi_{r}}^{\ast}[x(\tau)]e^{i\omega_{r}\tau} \non \\
&=& \left[f_{\Psi_{i}}[x(\tau)] \prod_{r = 1}^{n_{R}} f_{\Psi_{r}}^{\ast}[x(\tau)]\right] e^{i\Delta E\tau}.
\eqa 

Note that we have defined the Rindler space transition energy $\Delta E = \sum_{r=1}^{n_{R}} \omega_{r} - m_{i}$ to be the total energy difference between the final and initial Rindler states. We then find our acceleration-induced transition probability, Eq. (5), to be

\bqa
\mathcal{P} &=& \frac{G_{n}^{2}}{ \kappa} \iint d^{4}x d^{4}x' \sqrt{-g}\sqrt{-g'} \left|\bra{\prod_{j = 1}^{n_{R}}\Psi_{j}}   \hat{\Psi}_{i}(x) \prod_{r = 1}^{n_{R}}\hat{\Psi}_{r}(x) \ket{\Psi_{i}}\right|^{2}  \prod_{m = 1}^{n}G_{m}^{\pm}[x',x] \non \\
&=& \frac{G_{n}^{2}}{ \kappa} \iint d^{4}x d^{4}x' \sqrt{-g}\sqrt{-g'} \left|\left[f_{\Psi_{i}}[x(\tau)] \prod_{r = 1}^{n_{R}} f_{\Psi_{r}}^{\ast}[x(\tau)]\right] e^{i\Delta E\tau}\right|^{2}  \prod_{m = 1}^{n}G_{m}^{\pm}[x',x] \non \\
&=& \frac{G_{n}^{2}}{ \kappa} \iint d^{4}x d^{4}x' \sqrt{-g}\sqrt{-g'} \left|f_{\Psi_{i}}[x(\tau)] \prod_{r = 1}^{n_{R}} f_{\Psi_{r}}^{\ast}[x(\tau)] \right|^{2} e^{-i\Delta E(\tau'-\tau)} \prod_{m = 1}^{n}G_{m}^{\pm}[x',x].
\eqa

We can now use the overall normalization associated with the overlap of the spatial waveforms $\kappa$ [4,8] to simplify the above expression. In experimental settings the initial accelerated particle will be described by a wave packet with mode functions of the form $f_{\Psi}(x) \sim K_{i\omega/a}(\frac{m}{a}e^{a\xi}) g(\mathbf{x}_{\perp})$. The spatial distribution of the initial particle, in the directions perpendicular to the acceleration, is assumed to be finite, e.g. Gaussian, and is given by $g(\mathbf{x}_{\perp})$. With the mode functions properly normalized [13], $\kappa$ will be of order unity. Hence 

\bqa
\kappa = \iint d^{3}x d^{3}x' \sqrt{-g}\sqrt{-g'} \left|f_{\Psi_{i}}[x(\tau)] \prod_{r = 1}^{n_{R}} f_{\Psi_{r}}^{\ast}[x(\tau)] \right|^{2}.
\eqa

Having integrated over the spatial coordinates of the right Rindler wedge and noting that we still have to integrate over the proper times, it should be mentioned that the trajectories which characterize the Wightman functions must only depend on the proper time and have no spatial dependence. With this in mind, our transition probability becomes 

\bqa
\mathcal{P} &=& G_{n}^{2} \iint d\tau d\tau'  e^{-i\Delta E(\tau'-\tau)} \prod_{m = 1}^{n}G_{m}^{\pm}[x',x].
\eqa

Upon inspection of the above transition probability we find that we have now effectively reproduced the formalism the would be obtained if we had used an Unruh-DeWitt detector provided we identify the energy gap of the detector with the initial and final state Rindler energies $\Delta E$. To better incorporate the time dependence of our current analysis we will make the following change of variables to the rapidity variables $u'$ and $u$ defined by $u(\tau) =\int^{\tau} a(\tilde{\tau}) d\tilde{\tau}$. Hence

\bqa
d\tau d\tau' &=& du du' \frac{d\tau}{du}\frac{d\tau'}{du'} \non \\
&=& du du' \frac{1}{a}\frac{1}{a'}.
\eqa

We dropped the proper time dependencies in terms of the more compact notation $a'=a(\tau')$ which we also apply for all other variables. In terms of the rapidity, the transition probability then takes the following form

\bqa
\mathcal{P} &=& G_{n}^{2} \iint d\tau d\tau'  e^{-i\Delta E(\tau'-\tau)} \prod_{m = 1}^{n}G_{m}^{\pm}[x',x] \non \\
&=& G_{n}^{2}  \iint du du' \frac{1}{a}\frac{1}{a'} e^{-i\Delta E(\tau'-\tau)} \prod_{m = 1}^{n}G_{m}^{\pm}[x',x].
\eqa

Finally we decouple the integrals by formulating the new variables in terms of the difference $\xi = u'- u$ and average $\eta = \frac{u'+u}{2}$ rapidities. The inversions of these transformations are given by $u' = \xi/2 + \eta$ and $u = -\xi/2 + \eta$. With this new rapidity parametrization, the transition probability takes the similar form

\bqe
\mathcal{P} = G_{n}^{2} \iint d\eta d\xi \frac{1}{a'}\frac{1}{a} e^{-i\Delta E(\tau'-\tau)} \prod_{m = 1}^{n}G_{m}^{\pm}[x',x].
\eqe

For the sake of clarity we note that under this parametrization the primed and unprimed variables will all have dependencies such as $a' = a(\xi/2 +\eta)$ and $a = a(-\xi/2 + \eta)$. All components of the integrand depend on these variables in this manner. Moreover, we will eventually want to determine the transition rate, i.e. the probability per unit time $\Gamma = \frac{d \mathcal{P}}{d\tau}$. We shall choose the proper time parametrization $\tau_{\eta}$ that characterizes the rapidity variable $\eta$ via the definition $d \eta = a(\tau_{\eta}) d\tau_{\eta}$ for this purpose. Finally, using the notation $a(\tau_{\eta}) = a_{\eta}$ we find the transition rate to be

\bqa
\mathcal{P} &=& G_{n}^{2} \iint d\eta d\xi \frac{1}{a'a} e^{-i\Delta E(\tau'-\tau)} \prod_{m = 1}^{n}G_{m}^{\pm}[x',x] \non \\
\Rightarrow \Gamma &=& G_{n}^{2} \int d\xi \frac{a_{\eta}}{a'a} e^{-i\Delta E(\tau'-\tau)} \prod_{m = 1}^{n}G_{m}^{\pm}[x',x].
\eqa

We mention as well the ability to compute the differential transition probability per unit rapidity $\Gamma_{\eta} = \frac{d \mathcal{P}}{d\eta}$ that follows along with this derivation as well. To develop the integrand into a more useful form, we Taylor expand the proper time interval about the point $\xi = 0$. Recalling the form of the coordinate transformations used above, we find 

\bqa
\tau'-\tau &\sim & \tau_{0} + \left. \frac{d\tau'}{d u'}\frac{du'}{d \xi} \right|_{\xi = 0}\xi - \lb \tau_{0} + \left. \frac{d\tau}{d u}\frac{du}{d \xi} \right|_{\xi = 0}\xi \rb \non \\
&=&  \frac{1}{a_{\eta}}\frac{1}{2} \xi +  \frac{1}{a_{\eta}}\frac{1}{2} \xi \non \\ 
&=& \frac{\xi}{a_{\eta}}.
\eqa

Similarly, we expand the proper accelerations about the same point but we also disregard terms of order $j_{\eta}/a_{\eta}^2$ and higher. Hence 

\bqa
a'(\xi/2 +\eta) &\sim & a_{\eta} + \left. \frac{da'}{d\tau'}\frac{d\tau'}{du'}\frac{du'}{d\xi} \right|_{\xi = 0}\xi \non \\
& = & a_{\eta} + \left. J'\frac{1}{a'}\frac{1}{2} \right|_{\xi = 0} \xi \non \\
& = & a_{\eta} +  \frac{J_{\eta}}{a_{\eta}}\frac{1}{2}\xi \non \\
& = & a_{\eta}[1 +  \frac{J_{\eta}}{a_{\eta}^{2}}\frac{1}{2}\xi] \non \\
& \sim & a_{\eta}.
\eqa 

Similarly, we obtain $a(-\xi/2 +\eta) \sim a_{\eta}$ as well. We must also recall that all other components of the integrand must be expanded to the appropriate order. As such, we employ the notation $G_{m}^{\pm}[x',x]_{\eta}$ to imply the necessary Taylor expansion of the Wightman function. Moreover, now that acceleration is a function of one variable only, we will drop the $\eta$ subscript and define $a_{\eta} = a(\tau_{\eta}) \equiv a$. Thus our generalized response function takes the following form

\bqa
\Gamma &=& G_{n}^{2} \frac{1}{a}\int d\xi  e^{-i\Delta E\xi /a} \prod_{m = 1}^{n}G_{m}^{\pm}[x',x]_{\eta}.
\eqa

This expression determines the transition rate for a Rindler, i.e. accelerated, particle to decay into $n_{R}$ Rindler, i.e. accelerated, particles accompanied by the simultaneous emission of $n$ Minkowski, i.e. inertial, particles. This is the process that is expressed schematically in Eq. (1). A conceptual depiction of the above acceleration-induced process is also illustrated in Fig. 1 for clarity. To finalize this section we comment on the relationship between the method of field operators, used in this manuscript, with the alternative method of detectors [8]. The method of fields enables a more general analysis via the inclusion of an arbitrary number of Rindler particles of varying energy in the final state. The energy difference, as measured in the proper frame of the initial accelerated particle, is given by $\Delta E = \sum_{r=1}^{n_{R}} \omega_{r} - m_{i}$; see e.g. Eq. (6) and the subsequent discussion. That is, the difference between the sum of all final state Rindler particle energies and the initial accelerated particle's mass. To map the analysis to the method of detectors, all one needs to do is identify this energy difference with the energy gap of the detector, i.e. a two-level system. In short, the initial and final state energies of the Rindler particles, as measured in the proper frame of the initial accelerated particle, define the two energy levels of the detector. For a more in-depth discussion of this correspondence we refer the reader to Ref. [8]. In the next section we develop the time-dependent formalism that will be used to compute the Wightman functions and their subsequent Taylor expansion used in Eq. (16) above. 

\begin{figure}[H]
\centering  
\includegraphics[,scale=1]{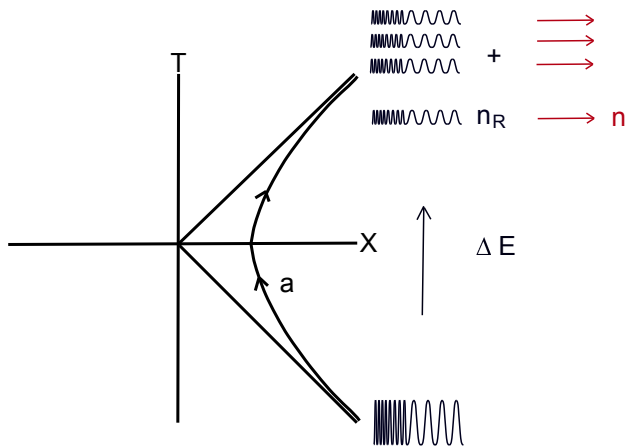}
\caption{A pictorial representation of the acceleration-induced transition from Eq. (16).}
\end{figure}

\section{Time-dependent spacetime trajectories}

Prior to evaluating the Wightman functions we will need the time-dependent spacetime trajectories the Minkowski fields propagate along. For the time-dependent proper acceleration $a(\tau)$, we recall the rapidity is defined by $u(\alpha) = \int^{\alpha} a(\tau) d\tau$. Using the Rindler chart to characterize the resultant accelerated motion, see e.g. [14], we have

\bqa
x(\tau) &=& \int^{\tau} d\alpha\sinh{[u(\alpha)]} + x_{0}\non \\
t(\tau) &=& \int^{\tau} d\alpha\cosh{[u(\alpha)]} + t_{0}.
\eqa

In order to compute these, and related, integrals we shall make use of a simple variant of the method of steepest descent. Here, we will maintain the appropriate expansion and again disregard terms of order $j/a^2$. As such, the method utilized can accommodate general acceleration profiles quite easily but depending on the system additional care must be taken if the acceleration goes to zero within the interval. Thus we consider the following change of variables and integration by parts [15],

\bqa
I &=& \int_{\tau}^{\tau'}d\alpha e^{\pm u(\alpha)} \non \\
&=& \int_{\tau}^{\tau'}\frac{1}{\pm a(\alpha)}\frac{d}{d\alpha}(e^{\pm u(\alpha)}) d\alpha \non \\
&=& \lbk \frac{1}{\pm a(\alpha)}e^{\pm u(\alpha)} \rbk_{\tau}^{\tau'} \pm\int_{\tau}^{\tau'}\frac{ j(\alpha)}{ a^{2}(\alpha)}e^{\pm u(\alpha)} d\alpha \non \\
&\approx &  \lbk \frac{1}{\pm a(\alpha)}e^{\pm u(\alpha)} \rbk_{\tau}^{\tau'}.
\eqa 

Note that we used the fact that $u'(\tau) = a(\tau)$ and $u''(\tau) = j(\tau)$. Moreover we see that we obtained a solution to the integral to zeroth order in $j/a^2$ as required to be consistent with the development of the generalized response function in the previous section. We then have the general form of the integral to be applied to our spacetime intervals. Hence

\bqa
\int_{\tau}^{\tau'}d\alpha e^{\pm u(\alpha)} &=& \frac{1}{\pm a(\tau')}e^{\pm u(\tau')} - \frac{1}{\pm a(\tau)}e^{\pm u(\tau)}. 
\eqa 

We now apply the above formula to compute all necessary spacetime quantities associated with our generalized hyperbolic trajectory. Considering first the spacelike interval $\Delta x = x' - x$, we find

\bqa
\Delta x  &=&  \int^{\tau'} d\alpha\sinh{[u(\alpha)]} + x_{0} - \lb \int^{\tau} d\alpha\sinh{[u(\alpha)]} + x_{0} \rb \non \\
&=& \int^{\tau'}_{\tau} d\alpha\sinh{[u(\alpha)]} \non \\
&=& \frac{1}{2}\int^{\tau'}_{\tau} d\alpha [e^{u(\alpha)} -e^{-u(\alpha)} ]\non \\
&=& \frac{1}{a(\tau')}\cosh{[ u(\tau')]} - \frac{1}{a(\tau)}\cosh{[ u(\tau)]}.
\eqa

Utilizing the same coordinate transformation, i.e. $u,u' \rightarrow \xi,\eta$ and Taylor expansion, i.e. $a'(\xi,\eta), a(\xi,\eta) \rightarrow a_{\eta} = a$ as in Sec. II, the above spacelike interval becomes

\bqa
\Delta x &=& \frac{1}{a(\tau')}\cosh{[ u(\tau')]} - \frac{1}{a(\tau)}\cosh{[ u(\tau)]} \non \\
&=& \frac{1}{a'(\eta,\xi)}\cosh{[ \xi /2+ \eta]} - \frac{1}{a(\eta, \xi)}\cosh{[ -\xi /2 +\eta]} \non \\
&=& \frac{1}{a}(\cosh{[ \xi /2+ \eta]} - \cosh{[ -\xi /2 +\eta]}) \non \\
&=& \frac{2}{a}\sinh{[ \xi /2]}\sinh{[ \eta]}.
\eqa

Note the use of the hyperbolic double angle formula to obtain the last line. Similarly, for the timelike interval $\Delta t = t' - t$ we find

\bqa
\Delta t  &=&  \int^{\tau'} d\alpha\cosh{[u(\alpha)]} + t_{0} - \lb \int^{\tau} d\alpha\cosh{[u(\alpha)]} + t_{0} \rb \non \\
&=& \int^{\tau'}_{\tau} d\alpha\cosh{[u(\alpha)]} \non \\
&=& \frac{1}{2}\int^{\tau'}_{\tau} d\alpha [e^{u(\alpha)} + e^{-u(\alpha)} ]\non \\
&=& \frac{1}{a(\tau')}\sinh{ [u(\tau')]} - \frac{1}{a(\tau)}\sinh{[ u(\tau)]}.
\eqa

The subsequent Taylor expansion to zeroth order in $j/a^{2}$ then yields

\bqa
\Delta t &=& \frac{1}{a(\tau')}\sinh{ [u(\tau')]} - \frac{1}{a(\tau)}\sinh{[ u(\tau)]} \non \\
&=& \frac{1}{a'(\eta,\xi)}\sinh{[ \xi /2+ \eta]} - \frac{1}{a(\eta, \xi)}\sinh{[ -\xi /2 +\eta]} \non \\
&=& \frac{1}{a}(\sinh{[ \xi /2+ \eta]} - \sinh{[ -\xi /2 +\eta]}) \non \\
&=& \frac{2}{a}\sinh{[ \xi /2]}\cosh{[ \eta]}.
\eqa

It should be noted that the addition of the complex regulator to the timelike interval can be added in without changing the computation. Next we compute the complex regulated square of the spacetime interval, $\Delta x^{2} - (\Delta t - i\epsilon)^{2}$, more commonly found in Wightman functions. We can evaluate this quantity by substitution of the previously computed spacelike and timelike intervals; however, in the interest of further developing the time-dependent formalism, we shall first carry out a few more manipulations to obtain a more general form. Thus

\bqa
\Delta x^{2} - ( \Delta t-i\epsilon )^{2} &=& \lbk x(\tau') - x(\tau)\rbk^{2} - \lbk t(\tau') - t(\tau) - i\epsilon \rbk^{2} \non \\
&=& \lbk \int^{\tau'} d\alpha\sinh{[u(\alpha)]} - \int^{\tau} d\alpha\sinh{[u(\alpha)]}\rbk^{2} - \lbk \int^{\tau'} d\alpha\cosh{[u(\alpha)]} - \int^{\tau} d\alpha\cosh{[u(\alpha)]} - i\epsilon\rbk^2 \non \\
&=& \lbk \int^{\tau'}_{\tau} d\alpha\sinh{[u(\alpha)]}\rbk^{2} - \lbk \int^{\tau'}_{\tau} d\alpha\cosh{[u(\alpha)]} - i\epsilon\rbk^2 \non \\
&=&  \iint^{\tau'}_{\tau} d\alpha d\beta \lb \sinh{[u(\alpha)]}\sinh{[u(\beta)]} - \cosh{[u(\alpha)]}\cosh{[u(\beta)]} \rb + 2i\epsilon\int^{\tau'}_{\tau} d\alpha \cosh{[u(\alpha)]} \non \\
&=&  -\iint^{\tau'}_{\tau} d\alpha d\beta \cosh{[u(\alpha)-u(\beta)]}  + 2i\epsilon \Delta t.
\eqa

Note in the last line that we used the hyperbolic double angle formula and rewrote the term on the complex regulator as $\Delta t$. Focusing on the integral, we break up the hyperbolic argument into exponentials to obtain a more convenient form. Hence

\bqa
\iint^{\tau'}_{\tau} d\alpha d\beta \cosh{(u(\alpha)-u(\beta))} &=& \frac{1}{2}\iint^{\tau'}_{\tau} d\alpha d\beta e^{u(\alpha) - u(\beta)} + e^{-(u(\alpha) - u(\beta))} \non \\
&=& \frac{1}{2}\int^{\tau'}_{\tau} d\alpha e^{u(\alpha)} \int^{\tau'}_{\tau}d\beta e^{-u(\beta)} + \frac{1}{2}\int^{\tau'}_{\tau} d\alpha e^{-u(\alpha)} \int^{\tau'}_{\tau}d\beta e^{u(\beta)} \non \\
&=& \int^{\tau'}_{\tau} d\alpha e^{u(\alpha)} \int^{\tau'}_{\tau}d\beta e^{-u(\beta)}.
\eqa

Note in the last line that we interchanged indices $\alpha \leftrightarrow \beta$ to combine the exponentials into one expression. We should also note a similar expression was obtained in [10]. Thus the square of the spacetime interval, along with its complex regulator, can be expressed in the following more compact form:

\bqe
\Delta x^{2} - ( \Delta t-i\epsilon )^{2} = \int^{\tau'}_{\tau} d\alpha e^{u(\alpha)} \int^{\tau'}_{\tau}d\beta e^{-u(\beta)} + 2i\epsilon \Delta t.
\eqe

In turning our task to evaluating the above integrals, we find

\bqa
\int^{\tau'}_{\tau} d\alpha e^{u(\alpha)} \int^{\tau'}_{\tau}d\beta e^{-u(\beta)} &=& - \lbk \frac{1}{ a(\tau')}e^{ u(\tau')} - \frac{1}{a(\tau)}e^{ u(\tau)} \rbk \lbk -\frac{1}{ a(\tau')}e^{- u(\tau')} + \frac{1}{ a(\tau)}e^{- u(\tau)}  \rbk \non \\
&=& \frac{-2}{a(\tau')a(\tau)}\cosh{[u(\tau')-u(\tau)]}+\frac{1}{a^{2}(\tau')}+\frac{1}{a^{2}(\tau)} \non \\
&=& \frac{-4}{a(\tau')a(\tau)}\sinh^{2}{[\frac{1}{2}(u(\tau')-u(\tau))]}-\frac{2}{a(\tau')a(\tau)}+\frac{1}{a^{2}(\tau')}+\frac{1}{a^{2}(\tau)} \non \\
&=& \frac{1}{a^{2}(\tau')a^{2}(\tau)}\lbk(a(\tau')-a(\tau))^{2} - 4a(\tau')a(\tau)\sinh^{2}{[\frac{1}{2}(u(\tau')-u(\tau))]} \rbk.
\eqa

Finally, performing the necessary change of variables and Taylor expanding the acceleration we obtain the final form of the $i\epsilon$ regularized square of the spacetime interval. Hence

\bqa
\Delta x^{2} - ( \Delta t-i\epsilon )^{2} &=& \frac{1}{a^{2}(\tau')a^{2}(\tau)}\lbk(a(\tau')-a(\tau))^{2} - 4a(\tau')a(\tau)\sinh^{2}{[\frac{1}{2}(u(\tau')-u(\tau))]}  \rbk + 2i\epsilon\Delta t \non \\
&=& \frac{1}{a'^{2}(\eta, \xi)a^{2}(\eta, \xi)}\lbk(a'(\eta, \xi)-a(\eta, \xi))^{2} - 4a'(\eta, \xi)a(\eta, \xi)\sinh^{2}{[\xi/2]} \rbk + 2i\epsilon\Delta t\non \\
&=& -\frac{4}{a^{2}}\sinh^{2}{[\xi/2]}  + i\epsilon\frac{4}{a}\sinh{[ \xi /2]}\cosh{[ \eta]} \non \\
&=& -\frac{4}{a^{2}}\sinh^{2}{[\xi/2]}  + i\epsilon\frac{8a}{a^2}\sinh{[ \xi /2]}\cosh{[ \xi /2]} \non \\
&=& -\frac{4}{a^{2}}\sinh^{2}{[\xi/2-ia\epsilon]} \non \\
&=& -\frac{4}{a^{2}}\sinh^{2}{[\xi/2-\sgn{(a)}i\epsilon]}.
\eqa

Note that, in the third to last line, we pulled the positive definite factor $\frac{2\cosh{[ \xi /2]}}{\cosh{[\eta]}}$ out of the $\epsilon$ and in the last line we absorbed the magnitude of the acceleration into the regulator. We keep the overall sign of the acceleration so as to not change the direction of the shift along the imaginary axis by changing the sign of $\epsilon$. To finalize the development of the components necessary to compute each of our Wightman functions, we also require the use of a Lorentz gamma to boost forward and backward between the proper and inertial lab frames. The Lorentz gamma can be computed by taking the derivative of the inertial time. We also parametrize our proper time here using the variable $\tau_{\eta}$ such that $u(\tau_{\eta}) = \eta$ as in the previous section. Recalling $dt = \gamma d\tau$ we have

\bqa
\frac{d}{d \tau_{\eta}}t &=& \frac{d}{d \tau_{\eta}} \int^{\tau_{\eta}}d\alpha \cosh{[u(\alpha)]} + t_{0} \non \\
\Rightarrow \gamma &=& \cosh{[\eta]}.
\eqa

In the next section we will utilize the above dictionary of formulas to evaluate the Wightman functions and its variants. The Wightman functions will then be used to compute the acceleration-induced transition rate, power emitted, and spectra of emitted particles. 

\section{The Wightman function and its variants}

In this section we compute the variants of the Wightman functions used for the computation of various observables. We will be working under the assumption that all Minkowski fields are massless. To evaluate the vacuum to vacuum two-point function [16], we use the canonically normalized field operator $\hat{\phi} = \int \frac{ d^{3}k}{(2 \pi)^{3/2}\sqrt{2\omega}} [ \hat{a}_{\mathbf{k}}e^{i(\mathbf{k}\cdot \mathbf{x} -\omega t)}  + \hat{a}_{\mathbf{k}}^{\dagger}e^{-i(\mathbf{k}\cdot \mathbf{x} -\omega t)}]$. Thus,

\bqa
G_{m}^{\pm}[x',x] &=& \bra{0}\hat{\phi}_{m}(x')\hat{\phi}_{m}(x)\ket{0} \nonumber \\
&=& \frac{1}{2(2 \pi)^{3}}\iint\frac{ d^{3}k'd^{3}k}{\sqrt{\omega'\omega}}\bra{0} \hat{a}_{\mathbf{k'}}\hat{a}_{\mathbf{k}}^{\dagger}e^{i(\mathbf{k'}\cdot \mathbf{x'} -\mathbf{k}\cdot \mathbf{x} -\omega' t'+\omega t)}\ket{0} \nonumber \\
&=& \frac{1}{2(2 \pi)^{3}}\iint\frac{ d^{3}k'd^{3}k}{\sqrt{\omega'\omega}}e^{i(\mathbf{k'}\cdot \mathbf{x'} -\mathbf{k}\cdot \mathbf{x} -\omega' t'+\omega t)}\delta(\mathbf{k'}-\mathbf{k}) \nonumber \\
&=& \frac{1}{2(2 \pi)^{3}}\int\frac{d^{3}k}{\omega}e^{i(\mathbf{k}\cdot \Delta\mathbf{x} -\omega \Delta t)} \non \\
&=& \frac{1}{2(2 \pi)^{3}}\int\frac{d^{3}k}{\omega}e^{-ik^{\m}\Delta x_{\m}}.
\eqa   

Since we are dealing with the emission of Minkowski particles by an accelerated field, it is advantageous to boost the momenta into the frame that is instantaneously at rest with the accelerated field. The integration measure is Lorentz invariant, i.e. $\frac{d^{3}k}{\omega} = \frac{d^{3}\tilde{k}}{\tilde{\omega}}$. When boosting the momenta forward into the accelerated fields' instantaneous rest frame via $k^{\m}\rightarrow \Lambda^{\m}_{\nu}k^{\nu} = \tilde{k}^{\m}$ , the Lorentz scalar nature of the exponent necessitates the boosting of the spacetime interval back via $\Delta x^{\m} \rightarrow (\Lambda^{-1})^{\m}_{\nu}\Delta x^{\nu} = \Delta \tilde{x}^{\m}$. It should be noted that each $\Delta \tilde{x}^{\m}$ is a proper quantity that has just been boosted to the same velocity but in the opposite direction and therefore all of the relativistic effects, e.g. length contraction, remain the same. Also, here and throughout, we will denote all proper quantities with a tilde. Therefore, our Wightman function, evaluated in the proper frame, is given by

\bqa
G_{m}^{\pm}[x',x] &\equiv& \frac{1}{2(2 \pi)^{3}}\int\frac{d^{3}\tilde{k}}{\tilde{\omega}} e^{i(\tilde{\mathbf{k}}\cdot \Delta\tilde{\mathbf{x}} -\tilde{\omega} \Delta \tilde{t})}.
\eqa

Integrals of this form are best evaluated in spherical coordinates. Without loss of generality we rotate the coordinate system until the momentum lies along the $z$-axis and then we enforce the condition that our Minkowski fields are massless, i.e. $\tilde{\omega} = \tilde{k}$. As such, the integral reduces to

\bqa
G_{m}^{\pm}[x',x] &=& \frac{1}{2(2 \pi)^{3}}\int\frac{d^{3}\tilde{k}}{\tilde{\omega}}  e^{i(\tilde{\mathbf{k}}\cdot \Delta\tilde{\mathbf{x}} -\tilde{\omega} \Delta \tilde{t})} \nonumber \\
&=& \frac{1}{2(2 \pi)^{3}}\int_{0}^{\infty}\int_{0}^{\pi}\int_{0}^{2\pi} d\tilde{\omega} d\tilde{\theta} d\tilde{\phi} \;\tilde{\omega}\sin{\tilde{\theta}}e^{i(\tilde{\omega}\Delta \tilde{x} \cos{\tilde{\theta}} -\tilde{\omega} \Delta \tilde{t})} \nonumber \\
&=& \frac{1}{2(2 \pi)^{2}}\int_{0}^{\infty}\int_{-1}^{1} d\tilde{\omega}d(\cos{\tilde{\theta}}) \;\tilde{\omega}e^{i(\tilde{\omega}\Delta \tilde{x} \cos{\tilde{\theta}} -\tilde{\omega} \Delta \tilde{t})} \nonumber \\
&=& \frac{1}{2(2 \pi)^{2}}\frac{1}{i \Delta \tilde{x}} \int_{0}^{\infty} d\tilde{\omega}\lbk e^{i\tilde{\omega}(\Delta \tilde{x} - \Delta \tilde{t})} - e^{-i\tilde{\omega}(\Delta \tilde{x} + \Delta \tilde{t})} \rbk.
\eqa

The above Wightman function will be used to compute the acceleration-induced transition rate. To properly evaluate the integral we require an infinitesimal shift along the imaginary time axis to regulate the oscillations at infinity. This is accomplished by letting $\Delta \tilde{t} \rightarrow \Delta \tilde{t} - i\epsilon$ where $\epsilon >0$. Thus, 

\bqa
G_{m}^{\pm}[x',x] &=& \frac{1}{2(2 \pi)^{2}}\frac{1}{i \Delta \tilde{x}} \int_{0}^{\infty} d\tilde{\omega} \lbk e^{i\tilde{\omega}(\Delta \tilde{x} - (\Delta \tilde{t}-i\epsilon))} - e^{-i\tilde{\omega}(\Delta \tilde{x} + (\Delta \tilde{t} -i\epsilon))} \rbk \nonumber \\
&=&-\frac{1}{2(2 \pi)^{2}}\frac{1}{i \Delta \tilde{x}}\lbk \frac{1}{i(\Delta \tilde{x} - (\Delta \tilde{t}-i\epsilon))} + \frac{1}{i(\Delta \tilde{x} + (\Delta \tilde{t}-i\epsilon))}  \rbk \nonumber \\
&=&\frac{1}{(2 \pi)^{2}} \frac{1}{\Delta \tilde{x}^{2} - (\Delta \tilde{t}-i\epsilon)^{2}} \non \\
&=&-\frac{1}{(2 \pi)^{2}} \frac{1}{\Delta \tilde{x}^{\m}\Delta \tilde{x}_{\m}} \non \\
&=&-\frac{1}{(2 \pi)^{2}} \frac{1}{\Delta x^{\m}\Delta x_{\m}}.
\eqa

Note in the last line we used the Lorentz invariance of the scalar product to boost the spacetime interval forward into the lab frame. This will facilitate later computations and also highlights the appropriate Lorentz invariance of the Wightman functions. It serves to also remember the presence of the complex regulator within the time component of the interval. Using the appropriately Taylor expanded spacetime interval derived in Eq. (28) of the previous section, we find the Taylor expanded Wightman function $G_{m}^{\pm}[x',x]_{\eta}$ to be

\bqa
G_{m}^{\pm}[x',x]_{\eta} &=&-\frac{1}{(2 \pi)^{2}} \frac{1}{\Delta x^{\m}\Delta x_{\m}} \non \\
&=&-\frac{1}{(2 \pi)^{2}} \frac{1}{\frac{4}{a^{2}}\sinh^{2}{[\xi/2-\sgn{(a)}i\epsilon]} } \non \\
&=&-\frac{a^{2}}{(4 \pi)^{2}} \frac{1}{\sinh^{2}{[\xi/2-\sgn{(a)}i\epsilon]}}.
\eqa

The Wightman function characterizes the probability for a particle to propagate along a given trajectory and is summed over all energies, i.e. integrated. Thus, by multiplying each probability by the energy (see e.g. [6]), and then integrating, we can effectively compute the average energy carried by the particle during the transition process, i.e. the power radiated. Denoting this energy weighted Wightman function as $\mathcal{G}_{m}^{\pm}[x',x]_{\eta}$, we then carry out its computation in a similar manner. Beginning with Eq. (32), we find 

\bqa
\mathcal{G}_{m}^{\pm}[x',x] &=& \frac{1}{2(2 \pi)^{2}}\frac{1}{i \Delta \tilde{x}} \int_{0}^{\infty} d\tilde{\omega} \tilde{\omega} \lbk e^{i\tilde{\omega}(\Delta \tilde{x} - (\Delta \tilde{t}-i\epsilon))} - e^{-i\tilde{\omega}(\Delta \tilde{x} + (\Delta \tilde{t} -i\epsilon))} \rbk \nonumber \\
&=&\frac{1}{2(2 \pi)^{2}}\frac{1}{i\Delta \tilde{x}}\lbk \frac{1}{[i(\Delta \tilde{x} - (\Delta \tilde{t}-i\epsilon))]^{2}} - \frac{1}{[i(\Delta \tilde{x} + (\Delta \tilde{t}-i\epsilon))]^{2}}  \rbk \nonumber \\
&=&-\frac{1}{2 i\pi^{2}} \frac{\Delta \tilde{t}-i\epsilon}{[\Delta \tilde{x}^{2} - (\Delta \tilde{t}-i\epsilon)^{2}]^{2}} \non \\
&=&-\frac{1}{2 i \pi^{2}} \frac{\Delta \tilde{t}-i\epsilon}{[\Delta \tilde{x}^{\m}\Delta \tilde{x}_{\m}]^{2}} \non \\
&=&-\frac{1}{2 i \pi^{2}} \frac{\Delta t/\gamma-i\epsilon}{[\Delta x^{\m}\Delta x_{\m}]^{2}}.
\eqa

Note that the presence of the proper frame timelike interval in the above quantity necessitated the use of the Lorentz gamma to boost it back to the lab frame via $\Delta \tilde{t} = \Delta \tau = \Delta t/\gamma$. Evaluation of this energy weighted Wightman function along the Taylor expanded hyperbolic trajectory yields

\bqa
\mathcal{G}_{m}^{\pm}[x',x]_{\eta} &=& -\frac{1}{2 i \pi^{2}} \frac{\Delta \tilde{t}-i\epsilon}{[\Delta x^{\m}\Delta x_{\m}]^{2}} \non \\
&=& -\frac{1}{2 i \pi^{2}} \frac{\frac{2}{a}\sinh{[ \xi /2]}-i\epsilon}{[-\frac{4}{a^{2}}\sinh^{2}{[\xi/2-\sgn{(a)}i\epsilon]}]^{2}} \non \\
&=& -\frac{a^{4}}{ i(4 \pi)^{2}} \frac{\sinh{[ \xi /2]}-i\frac{a}{2}\epsilon}{\sinh^{4}{[\xi/2-\sgn{(a)}i\epsilon]}} \non \\
&=& -\frac{a^{3}}{ i(4 \pi)^{2}} \frac{\sinh{[ \xi /2]}-i\sgn{(a)}\epsilon \cosh{[\xi/2]}}{\sinh^{4}{[\xi/2-\sgn{(a)}i\epsilon]}} \non \\
&=& -\frac{a^{3}}{ i(4 \pi)^{2}} \frac{1}{\sinh^{3}{[\xi/2-\sgn{(a)}i\epsilon]}}.
\eqa

Again we pulled a positive definite factor $2\cosh{[\xi/2]}$ out of the $\epsilon$. We also absorbed the magnitude of the acceleration into the regulator and then recombined the numerator into the hyperbolic sine since it was the Taylor expansion about small $\epsilon$. Finally, examining how the propagation, and thus emission probability, changes as we vary the energy enables us to compute the spectra. Taking the derivative of the Wightman function, Eq. (32), yields

\bqa
\frac{d}{d\tilde{\omega}} G_{m}^{\pm}[x',x]&=& \frac{1}{2(2 \pi)^{2}}\frac{1}{i\Delta \tilde{x}} \lbk e^{i\tilde{\omega}(\Delta \tilde{x} - \Delta \tilde{t})} - e^{-i\tilde{\omega}(\Delta \tilde{x} + \Delta \tilde{t})}\rbk.
\eqa

Substitution of the relevant trajectories, along with the appropriate expansion, yields

\bqa
\frac{d}{d\tilde{\omega}} G_{m}^{\pm}[x',x]_{\eta} &=& \frac{1}{2(2 \pi)^{2}}\frac{1}{i\Delta \tilde{x}} \lbk e^{i\tilde{\omega}(\Delta \tilde{x} - \Delta \tilde{t})} - e^{-i\tilde{\omega}(\Delta \tilde{x} + \Delta \tilde{t})}\rbk \non \\
&=& \frac{1}{(2 \pi)^{2}}\frac{e^{-i\tilde{\omega}\Delta \tilde{t}}}{\Delta \tilde{x}} \sin{(\tilde{\omega}\Delta \tilde{x})} \non \\
&=& \frac{1}{(2 \pi)^{2}}\frac{e^{-i\tilde{\omega}\Delta t/\gamma}}{\gamma \Delta x} \sin{(\tilde{\omega} \gamma \Delta x)} \non \\
&=& \frac{a}{(2 \pi)^{2}}\frac{e^{-i\tilde{\omega}\frac{2}{a}\sinh{(\xi/2)}}}{\sinh{(\xi/2)}\sinh{(2\eta)}} \sin{\lb \frac{\tilde{\omega}}{a} \sinh{(\xi/2)}\sinh{(2\eta)} \rb } \non \\
&\approx & \frac{\tilde{\omega}}{(2 \pi)^{2}}e^{-i\tilde{\omega}\xi/a}.
\eqa

In the last line we expanded the arguments about small $\xi$. Note that we have kept, to first order, a dependence in the phase so we still encode the dynamics. Now that we have our catalog of the Wightman function and its variants, the remaining sections will be devoted to the computation of the relevant observables of the theory. 
 
\section{Transition Rate}

The generalized response function developed previously will now be used to compute the acceleration-induced transition rate. This will enable us to analyze the decay of unstable particles as well as the excitation of stable particles into states of higher energy. We begin by recalling the functional form of the response function, Eq. (16),

\bqa
\Gamma &=& G_{n}^{2} \frac{1}{a} \int d\xi  e^{-i\Delta E\xi /a} \prod_{m = 1}^{n}G_{m}^{\pm}[x',x]_{\eta}.
\eqa

To compute the transition rate we use all $n$ of the Wightman functions in the above product over our final state particles. Recalling the form of the Wightman functions, Eq. (34), we find that the transition rate of $n$-particle multiplicity simplifies to

\bqa
\Gamma &=& G_{n}^{2}\frac{1}{a}\int d\xi  e^{-i\Delta E\xi /a} \prod_{m = 1}^{n}G_{m}^{\pm}[x',x]_{\eta} \non \\
&=& G_{n}^{2} \frac{1}{a} \int d\xi  e^{-i\Delta E\xi /a}\lbk-\frac{a^{2}}{(4 \pi)^{2}} \frac{1}{\sinh^{2}{[\xi/2-\sgn{(a)}i\epsilon]}}\rbk^{n} \non \\
&=& G_{n}^{2} \lb\frac{ia}{4 \pi}\rb^{2n}\frac{\sgn{(a)}}{|a|} \int d\xi \frac{e^{-i\Delta E\xi /a}}{\sinh^{2n}{[\xi/2-\sgn{(a)}i\epsilon]}}.
\eqa

At this point we must focus the integral. Specifically we must note how the integration depends on the sign of the acceleration. By letting $\xi \rightarrow  \sgn{(a)}\xi$ we see the integration is independent of the sign of the acceleration. As such we set $\sgn{(a)} = 1$, leaving the transition rate dependent only on the magnitude of the acceleration $|a|$. Hence,

\bqa
\Gamma &=& G_{n}^{2} \lb\frac{ia}{4 \pi}\rb^{2n}\frac{1}{|a|} \int d\xi  \frac{e^{-i\Delta E\xi /|a|}}{\sinh^{2n}{[\xi/2-i\epsilon]}}.
\eqa
   
We now see that the removal of the $i\epsilon$ yields a singularity structure with poles of order $2n$ at $\xi =  2 \pi i \sigma$ with the integer $\sigma \geq 0$. Now that we know the pole structure, we rid ourselves of the complex regulator. Rewriting the denominator of the integrand in exponential form will yield a more useful form. Hence,

\bqa
\Gamma &=& G_{n}^{2} \lb\frac{ia}{4 \pi}\rb^{2n}\frac{1}{|a|} \int d\xi  \frac{e^{-i\Delta E\xi /|a|}}{\sinh^{2n}{[\xi/2-i\epsilon]}} \non \\
&=& G_{n}^{2} \lb\frac{ia}{2 \pi}\rb^{2n}\frac{1}{|a|} \int d\xi  \frac{e^{-i\Delta E\xi /|a|}}{[e^{\xi/2} - e^{-\xi/2}]^{2n}}. 
\eqa

Employing the more convenient change of variables $w = e^{\xi}$ we find

\bqa
\Gamma &=& G_{n}^{2} \lb\frac{ia}{2 \pi}\rb^{2n}\frac{1}{|a|} \int d\xi  \frac{e^{-i\Delta E\xi /|a|}}{[e^{\xi/2} - e^{-\xi/2}]^{2n}} \non \\
 &=& G_{n}^{2} \lb\frac{ia}{2 \pi}\rb^{2n}\frac{1}{|a|} \int dw  \frac{w^{-i\Delta E\xi /|a|+n-1}}{[w-1]^{2n}}.
\eqa

The above expression can be integrated using the residue theorem. The integral is of the form $\frac{w^{-i\beta+\gamma}}{[w-1]^{\delta}}$ with the appropriate identification for $\beta$, $\gamma$, and $\delta$. Also note the added conditions that $\gamma$ is an integer and $\delta$ is an even integer. Evaluation of this generalized integral yields

\bqa
\int dw \frac{w^{-i\beta+\gamma}}{[w-1]^{\delta}} &=& \frac{2 \pi i}{(\delta-1)!}\sum_{\sigma = 0}^{\infty}\frac{d^{\delta -1}}{dw^{\delta -1}}\lbk\lbk w-1 \rbk^{\delta}\frac{w^{-i\beta+\gamma}}{\lbk w-1 \rbk^{\delta}}\rbk_{w = e^{i2\pi \sigma}} \non \\
&=& \frac{2 \pi i}{(\delta -1)!}\frac{\Gamma(1-i\beta+\gamma)}{\Gamma(-i\beta + \gamma -\delta +2)}\sum_{\sigma = 0}^{\infty}\lbk w^{-i\beta+\gamma - \delta +1}\rbk_{w = e^{i2\pi \sigma}} \nonumber \\
&=& \frac{2 \pi i}{(\delta -1)!}\frac{\Gamma(1-i\beta+\gamma)}{\Gamma(-i\beta + \gamma -\delta +2)}\sum_{\sigma = 0}^{\infty} e^{2\pi \sigma \beta + i2\pi \sigma(\gamma - \delta)} \nonumber \\
&=& \frac{2 \pi i}{(\delta -1)!}\frac{\Gamma(1-i\beta+\gamma)}{\Gamma(-i\beta + \gamma -\delta +2)} \frac{1}{1 - e^{2\pi\beta}} \non \\
&=& \frac{2 \pi i}{(\delta -1)!}\frac{\Gamma(i\beta - \gamma + \delta -1)}{\Gamma(i\beta - \gamma )} \frac{1}{e^{2\pi\beta} - 1}.
\eqa

Note that we used the identity $\Gamma(z)\Gamma(1-z) = \pi/\sin(\pi z)$, along with the properties of $\gamma$ and $\delta$, in the last line. The sum in the above equation converges for $e^{2\pi\beta}<1$ which is true for negative $\beta$. This corresponds to $\Delta E<0$, i.e. decays. In order to evaluate the sum we may assume $\beta$ to be negative to yield the closed form expression of the convergent sum. We can then relax the condition for $\Delta E > 0$ and we note that, in the zero acceleration limit, the transition rate appropriately diverges. This merely illustrates the fact that inertially stable particles have an infinite lifetime. Utilizing the above derived formula, we evaluate the integral in Eq. (43) to be

\bqa
\int dw  \frac{w^{-i\Delta E\xi /|a|+n-1}}{[w-1]^{2n}} = \frac{2 \pi i}{(2n -1)!}\frac{\Gamma(i\Delta E/|a| + n)}{\Gamma(i\Delta E/|a| +1-n )}\frac{1}{e^{2\pi\Delta E/|a|} - 1}.
\eqa

The transition rate is then given by

\bqe
\Gamma(\Delta E,\eta) = G_{n}^{2} \lb \frac{ia}{2 \pi}\rb^{2n} \frac{1}{a}\frac{2 \pi i}{(2n -1)!}\frac{\Gamma(i\Delta E/|a| + n)}{\Gamma(i\Delta E/|a| +1-n )} \frac{1}{e^{2\pi\Delta E/|a|} - 1}.
\eqe

We will take one final step in simplifying the above gamma functions using the Pochhammer symbol for rising factorials $(x)^{a} = \frac{\Gamma{(x+a)}}{\Gamma{(x)}} = \prod_{j = 0}^{a-1}(x+j)$. This serves to better present the resultant polynomials of multiplicity. Applying the Pochhammer identity to the above expression, along with the identifications $x = i\Delta E/|a| +1-n$ and $a = 2n-1$, yields

\bqa
\frac{\Gamma(i\Delta E/|a| + n)}{\Gamma(i\Delta E/|a| +1-n )} &=& \prod_{j = 0}^{2n-2}(i\Delta E/|a| +1-n+j) \non \\  
&=& \prod_{k = -(n-1)}^{n-1}(i\Delta E/|a| + k) \non \\  
&=& \frac{|a|}{i\Delta E}\prod_{k = 0}^{n-1}(i\Delta E/|a| + k)(i\Delta E/|a| - k) \non \\ 
&=& (-1)^{n}\frac{|a|}{i\Delta E}\prod_{k = 0}^{n-1}\lbk (\Delta E/a)^{2}  +k^{2}\rbk \non \\  
&=&  \lb \frac{i\Delta E}{|a|} \rb^{2n-1}\prod_{k = 0}^{n-1}\lbk 1  + k^{2} \lb \frac{a}{\Delta E}\rb^{2} \rbk .  
\eqa

Utilizing the above expression, along with the double factorial identity $(2x)!! = 2^{x}(x)!$, we finalize the computation of the acceleration-induced transition rate. Thus

\bqa
\Gamma(\Delta E,\eta) = G_{n}^{2}\lb\frac{\Delta E}{\pi} \rb^{2n-1} \frac{1}{(4n -2)!!} \prod_{k = 0}^{n-1}\lbk 1  + k^{2} \lb \frac{ a}{\Delta E} \rb^{2} \rbk  \frac{1}{e^{2\pi\Delta E/|a|} - 1}.
\eqa

The above transition rates are characterized by a bosonic distribution along with the integer indexed polynomial of multiplicity. Normalizing the rates to the multiplicity-dependent coupling via $\tilde{\Gamma}=\Gamma/G_{n}^2$, we explicitly compute the normalized rates for the first five multiplicities.  Hence

\bqa
\tilde{\Gamma}_{1}(\Delta E, a) &=& \frac{\Delta E}{2 \pi} \frac{1}{e^{2\pi\Delta E/|a|}-1} \non \\
\tilde{\Gamma}_{2}(\Delta E, a) &=& \frac{\Delta E^{3}}{48 \pi^{3}} \frac{1+\lb \frac{a}{\Delta E}  \rb^{2}}{e^{2\pi\Delta E/|a|}-1} \non \\
\tilde{\Gamma}_{3}(\Delta E, a) &=& \frac{\Delta E^{5}}{3840 \pi^{5}} \frac{1+ 5 \lb \frac{a}{\Delta E}  \rb^{2} + 4 \lb \frac{a}{\Delta E}  \rb^{4}}{e^{2\pi\Delta E/|a|}-1} \non \\
\tilde{\Gamma}_{4}(\Delta E, a) &=& \frac{\Delta E^{7}}{645120 \pi^{7}} \frac{1+ 14 \lb \frac{a}{\Delta E}  \rb^{2} + 49 \lb \frac{a}{\Delta E}  \rb^{4} + 36 \lb \frac{a}{\Delta E}  \rb^{6}}{e^{2\pi\Delta E/|a|}-1} \non \\
\tilde{\Gamma}_{5}(\Delta E, a) &=& \frac{\Delta E^{9}}{185794560 \pi^{9}} \frac{1+ 30 \lb \frac{a}{\Delta E}  \rb^{2} + 273 \lb \frac{a}{\Delta E}  \rb^{4} + 820 \lb \frac{a}{\Delta E}  \rb^{6} + 576 \lb \frac{a}{\Delta E}  \rb^{8} }{e^{2\pi\Delta E/|a|}-1}.
\eqa 

The $n=1$ case is the standard result for computing the transition rate of an Unruh-DeWitt detector. We should also mention that the acceleration-dependent lifetime is easily computed via $\tau =1/\Gamma$. We plot  (see Figs. 2 and 3) the transition rates for acceleration-induced decay and excitation respectively. The fact that the excitation rate rapidly goes to zero, in the inertial limit, reflects the infinite lifetime for such processes. Moreover, it should be noted that there is an acceleration-dependent crossover scale where one multiplicity dominates the transition process. The next section deals specifically with these crossovers.  

\begin{figure}[H]
\centering  
\includegraphics[,scale=.6]{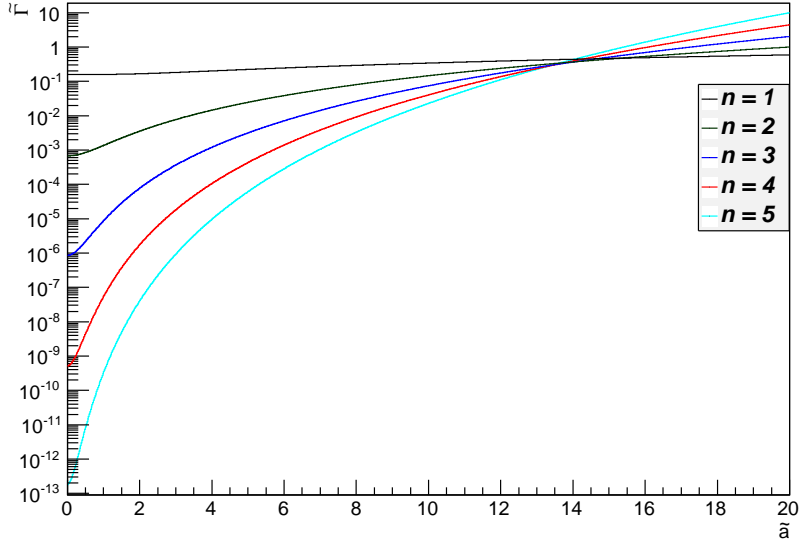}
\caption{The normalized transition rates, Eq. (49), with $\tilde{a} = |a|/\Delta E$ and $\Delta E=-1$.}
\end{figure} 

\begin{figure}[H]
\centering  
\includegraphics[,scale=.6]{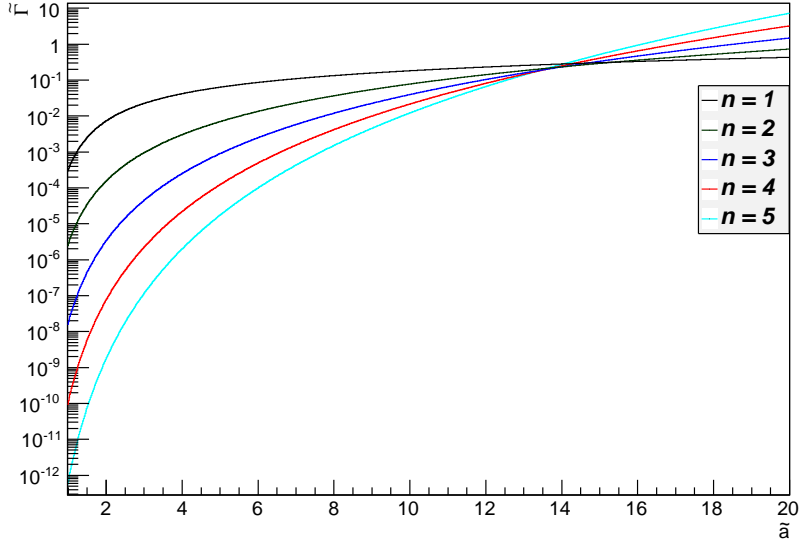}
\caption{The normalized transition rates, Eq. (49), with $\tilde{a} = |a|/\Delta E$ and $\Delta E=1$.}
\end{figure}

\section{Multiplicity}

In this section we compute the dominant multiplicity as a function of acceleration. Specifically, for final states which contain either $n$ or $m$ Minkowski particles, we ask when the transition rate for an $n$-particle final state is greater than an $m$-particle final state. This will enable us to characterize the acceleration scale which selects a specific transition when there are multiple decay pathways. Let us begin by computing the inertial limit of the decay rates. This enables us to fix the couplings $G^{2}_{n}$ to that of the inertial decay rate $\lambda_{n}$. For decays we have $\Delta E<0$ and taking the limit $a \rightarrow 0$ of the transition rate, Eq. (48), yields

\bqa
\lambda_{n} &=& G^{2}_{n} \lb \frac{\Delta E}{\pi} \rb^{2n - 1} \frac{1}{(4n - 2)!!} \non  \\
\Rightarrow G^{2}_{n} &=& \lambda_{n} \lb \frac{\pi}{\Delta E} \rb^{2n - 1} (4n - 2)!!.
\eqa

In terms of the dimensionless acceleration $\tilde{a} = |a|/\Delta E$, the crossover scale, i.e. when $\Gamma_{n} = \Gamma_{m}$, can then be computed in terms of the inertial rates. Hence

\bqa
\Gamma_{n}  &=& \Gamma_{m}  \non \\
\lambda_{n} \prod_{k = 0}^{n-1} \lbk 1+k^{2}\tilde{a}^{2}\rbk &=& \lambda_{m}\prod_{j = 0}^{m-1} \lbk 1+j^{2}\tilde{a}^{2}\rbk.
\eqa

The ratio of the inertial decay rates $\lambda_{m}/\lambda_{n}$ is equivalent to the ratio of the branching fractions $Br_{m}/Br_{n} \equiv g$ of each decay pathway [17]. Assuming $n>m$ we find

\bqa
\lambda_{n} \prod_{k = 0}^{n-1} \lbk 1+k^{2}\tilde{a}^{2}\rbk &=& \lambda_{m} \prod_{j = 0}^{m-1} \lbk 1+j^{2}\tilde{a}^{2}\rbk \non \\
 \prod_{k = 0}^{n-1} \lbk 1+k^{2}\tilde{a}^{2}\rbk &=& g \prod_{j = 0}^{m -1} \lbk 1+j^{2}\tilde{a}^{2}\rbk \non \\
 \prod_{k = m}^{n - 1} \lbk 1+k^{2}\tilde{a}^{2}\rbk &=& g.
\eqa

The above equation defines the acceleration scale at which an $n$-particle multiplicity final state will dominate the $m$-particle multiplicity final state. We now examine in more detail the cases when $\tilde{a}k\gg 1$, $\tilde{a}k\ll 1$, $n= m + 1$, and $n = m + 2$. For the case of large acceleration, i.e. $\tilde{a}k \gg 1$, we find

\bqa
g &=& \prod_{k = m}^{n - 1} \lbk 1+k^{2}\tilde{a}^{2}\rbk \non \\
g &=&\tilde{a}^{2(n-m)}\prod_{k = m}^{n - 1}  k^{2}  \non \\
g &=&\tilde{a}^{2(n-m)}\lbk \frac{(n - 1)!}{(m - 1)!}\rbk^{2}  \non \\
\Rightarrow \tilde{a}_{\uparrow} &=& \lbk g \lbk \frac{(m - 1)!}{(n - 1)!} \rbk^{2}\rbk^{1/(2(n-m))}.
\eqa

In the case of small acceleration, i.e. $\tilde{a}k \ll 1$, we make use of the properties of logarithms to simplify the computation. Hence

\bqa
g &=& \prod_{k = m}^{n - 1} \lbk 1+k^{2}\tilde{a}^{2}\rbk \non \\
\ln{(g)} &=& \ln{\lb \prod_{k = m}^{n - 1} \lbk 1 +k^{2}\tilde{a}^{2}\rbk \rb }  \non \\
\ln{(g)} &=& \sum_{k = m}^{n - 1} \ln{ \lbk 1 +k^{2}\tilde{a}^{2}\rbk}   \non \\
\ln{(g)} &=& \sum_{k = m}^{n - 1} k^{2}\tilde{a}^{2} \non \\
g &=& 1+ \tilde{a}^{2}\sum_{k = m}^{n - 1} k^{2}
\eqa

The last line followed from Taylor expanding the resultant exponential. The sum of squares evaluates to $\sum_{k = m}^{n - 1} k^{2} = \frac{1}{6}(n - m) -\frac{1}{2}(n^{2} - m^{2})+\frac{1}{3}(n^{3} - m^{3})$. Thus the acceleration scale for multiplicity transitions at low acceleration is given by

\bqa
\tilde{a}_{\downarrow} &=& \sqrt{\frac{g-1}{\frac{1}{6}(n - m) -\frac{1}{2}(n^{2} - m^{2})+\frac{1}{3}(n^{3} - m^{3})}}.
\eqa

It should be noted that the above formula is most applicable for two decay pathways with nearly identical branching fractions, i.e. $g\sim 1$. The case when $n-m = 1$ can be used to characterize the emission of an extra photon in the decay process. We can trivially solve for the acceleration in this case. Hence

\bqa
\tilde{a}_{1} &=& \sqrt{\frac{g-1}{(n-1)^{2}}}.
\eqa

The case when $n-m = 2$ can characterize the emission of an additional particle-antiparticle pair during the decay process. This case can also be solved exactly. Hence

\bqa
g &=&  \prod_{k = n-2}^{n} \lbk 1+k^{2}\tilde{a}^{2}\rbk \non \\
g &=& \lbk 1+(n-1)^{2}\tilde{a}^{2}\rbk \lbk 1+(n-2)^{2}\tilde{a}^{2}\rbk  \non \\
\Rightarrow \tilde{a}_{2} &=& \lbk \frac{\sqrt{ [(n-1)^{2} + (n-2)^{2}]^{2}+4(g-1)(n-1)^{2}(n-2)^{2}    } - [(n-1)^{2} + (n-2)^{2}] }{2(n-1)^{2}(n-2)^{2}}\rbk^{1/2}.
\eqa

The above acceleration scales can be used to fine-tune a system to select a preferred decay pathway and also, if the acceleration of the system is known, predict the relevant branching fractions of the system under study. In the next section we shall analyze the power radiated away by an accelerated particle.

\section{Power Radiated}

In order to compute the power emitted by the $i$th particle $\mathcal{S}_{i}$ we use the energy weighted Wightman function $\mathcal{G}^{\pm}$ for that particle in the generalized transition rate from Eq. (16). Thus

\bqa
\Gamma &=& G_{n}^{2} \frac{1}{a} \int d\xi  e^{-i\Delta E\xi /a} \prod_{m = 1}^{n}G_{m}^{\pm}[x',x]_{\eta} \non \\
\Rightarrow \mathcal{S}_{i} &=& G_{n}^{2} \frac{1}{a} \int d\xi  e^{-i\Delta E\xi /a}\mathcal{G}_{i}^{\pm}[x',x]_{\eta}\prod_{m \neq i}^{n-1}G_{m}^{\pm}[x',x]_{\eta}.
\eqa

Note that we separated out the energy weighted Wightman function of the $i$th particle. Recalling the explicit forms of the appropriate Wightman functions, Eqs. (34) and (36), the power radiated away by the $i$th particle simplifies to

\bqa
\mathcal{S}_{i} &=& G_{n}^{2} \frac{1}{a} \int d\xi  e^{-i\Delta E\xi /a}\mathcal{G}_{i}^{\pm}[x',x]_{\eta}\prod_{m \neq i}^{n-1}G_{m}^{\pm}[x',x]_{\eta} \non \\
&=& G_{n}^{2} \frac{1}{a} \int d\xi  e^{-i\Delta E\xi /a}\lbk - \frac{a^{3}}{ i(4 \pi)^{2}} \frac{1}{\sinh^{3}{[\xi/2-\sgn{(a)}i\epsilon]}} \rbk \lbk   -\frac{a^{2}}{(4 \pi)^{2}} \frac{1}{\sinh^{2}{[\xi/2-\sgn{(a)}i\epsilon]}} \rbk^{n-1} \non \\
&=& -G_{n}^{2} 4\pi \lb \frac{ia}{4 \pi} \rb^{2n+1} \frac{1}{a} \int d\xi   \frac{e^{-i\Delta E\xi /a}}{\sinh^{2n+1}{[\xi/2-\sgn{(a)}i\epsilon]}}.
\eqa

Again, with the same rescaling $\xi \rightarrow \sgn{(a)} \xi$, the integral is invariant under the change of sign of the acceleration. Thus we let $\sgn{(a)} = 1$ as before. Hence 

\bqa
\mathcal{S}_{i} &=& -G_{n}^{2} 4 \pi\lb \frac{i |a|}{4 \pi} \rb^{2n+1}  \frac{1}{|a|} \int d\xi  \frac{e^{-i\Delta E \xi /|a|}}{\sinh^{2n+1}{[\xi/2-i\epsilon]}}.
\eqa

Note that we find a pole structure similar to the transition rate in the absence of the regulator, this time with poles of order $2n+1$ at $\xi = 2 \pi i \sigma$ with integer $\sigma \geq 0$. Using the same computational machinery as before, we employ the change of variables as the previous section, $w = e^{\xi}$. Thus

\bqa
\mathcal{S}_{i} &=& -G_{n}^{2} 4\pi\lb \frac{i |a|}{4 \pi} \rb^{2n+1}  \frac{1}{|a|} \int d\xi  \frac{e^{-i\Delta E \xi /|a|}}{\sinh^{2n+1}{[\xi/2-i\epsilon]}} \non \\
&=& -G_{n}^{2} 4\pi\lb \frac{i |a|}{2 \pi} \rb^{2n+1}  \frac{1}{|a|} \int d\xi  \frac{e^{-i\Delta E \xi /|a|}}{[e^{\xi/2}-e^{-\xi/2}]^{2n+1}} \non \\
&=&-G_{n}^{2} 4\pi\lb \frac{i |a|}{2 \pi} \rb^{2n+1}  \frac{1}{|a|} \int dw  \frac{w^{-i\Delta E /|a| +n -1/2}}{[w-1]^{2n+1}}.
\eqa

Examining the above integral in a similar manner as in the last section we note that it is still of the form $\frac{w^{-i\beta+\gamma}}{[w-1]^{\delta}}$ but with odd integer $\delta$ and with $\gamma$ being an odd integer multiple of $1/2$. Let us also note that $\gamma - \delta$ will also be an odd integer multiple of $1/2$. The effect of this will be an alternating sign in the sum over the residues leading to a different thermal distribution. Hence

\bqa
\int dw \frac{w^{-i\beta+\gamma}}{[w-1]^{\delta}} &=& \frac{2 \pi i}{(\delta-1)!}\sum_{\sigma = 0}^{\infty}\frac{d^{\delta -1}}{dw^{\delta -1}}\lbk\lbk w-1 \rbk^{\delta}\frac{w^{-i\beta+\gamma}}{\lbk w-1 \rbk^{\delta}}\rbk_{w = e^{i2\pi \sigma}} \non \\
&=& \frac{2 \pi i}{(\delta -1)!}\frac{\Gamma(1-i\beta+\gamma)}{\Gamma(-i\beta + \gamma -\delta +2)}\sum_{\sigma = 0}^{\infty}\lbk w^{-i\beta+\gamma - \delta +1}\rbk_{w = e^{i2\pi \sigma}} \nonumber \\
&=& \frac{2 \pi i}{(\delta -1)!}\frac{\Gamma(1-i\beta+\gamma)}{\Gamma(-i\beta + \gamma -\delta +2)}\sum_{\sigma = 0}^{\infty} e^{2\pi \sigma\beta + i2\pi \sigma(\gamma - \delta)} \nonumber \\
&=& \frac{2 \pi i}{(\delta -1)!}\frac{\Gamma(1-i\beta+\gamma)}{\Gamma(-i\beta + \gamma -\delta +2)} \frac{1}{e^{2\pi\beta} +1 } \non \\
&=& \frac{2 \pi i}{(\delta -1)!}\frac{\Gamma(i\beta - \gamma + \delta -1)}{\Gamma(i\beta - \gamma )} \frac{1}{e^{2\pi\beta} + 1}.
\eqa

Again we used the properties of $\delta$ and $\gamma$ to manipulate the gamma functions in the last line. Then, utilizing the above formula, the integral in Eq. (61) yields

\bqe
\int dw  \frac{w^{-i\Delta E /|a| +n -1/2}}{[w-1]^{2n+1}} = \frac{2 \pi i}{(2n)!}\frac{\Gamma(i\Delta E/|a|+n+1/2)}{\Gamma(i\Delta E/|a| +1/2 - n)} \frac{1}{e^{2\pi\Delta E/|a|} + 1}
\eqe

The expression for the power radiated by the $i$th particle is then given by

\bqa
\mathcal{S}_{i} &=& -G_{n}^{2} 4\pi\lb \frac{i |a|}{2 \pi} \rb^{2n+1}  \frac{1}{|a|}\frac{2 \pi i}{(2n)!}\frac{\Gamma(i\Delta E/|a|+n+1/2)}{\Gamma(i\Delta E/|a| +1/2 - n)} \frac{1}{e^{2\pi\Delta E/|a|} + 1}.
\eqa

We again use the Pochhammer identity $\frac{\Gamma{(x+a)}}{\Gamma{(x)}} = \prod_{j = 0}^{a-1}(x+j)$ with the identifications $x = i\Delta E/|a| +1/2 - n$ and $a =2n$. As such, the above gamma functions yield a cleaner expression. Thus

\bqa
\frac{\Gamma(i\Delta E/|a|+n+1/2)}{\Gamma(i\Delta E/|a| +1/2 - n)} &=& \prod_{j = 0}^{2n-1}(i\Delta E/|a| + 1/2-n+ j) \non \\
&=&  \prod_{\ell_{odd} = -(2n-1)}^{2n-1}(i\Delta E/|a| + \ell /2) \non \\
&=&  \prod_{\ell_{odd} = 1}^{2n-1}(i\Delta E/|a| + \ell /2)(i\Delta E/|a| - \ell /2) \non \\
&=&  \lb \frac{i\Delta E}{|a|} \rb^{2n}\prod_{\ell_{odd} = 1}^{2n-1}\lbk 1 + (\ell /2)^{2} \lb \frac{a}{\Delta E} \rb^{2}\rbk \non \\
&=&  \lb \frac{i\Delta E}{|a|} \rb^{2n}\prod_{k = 0}^{n-1}\lbk 1 + \lb \frac{2k+1}{2} \rb^{2} \lb \frac{a}{\Delta E} \rb^{2}\rbk.
\eqa

Using the same double factorial identity $(2x)!! = 2^{x}(x)!$, we find that the power radiated by the $i$th particle is given by

\bqe
\mathcal{S}_{i} = G^{2}_{n}4\pi \lb \frac{\Delta E}{\pi} \rb^{2n} \frac{1}{(4n)!!}  \prod_{k = 0}^{n-1}\lbk 1 + \lb \frac{2k+1}{2} \rb^{2} \lb \frac{a}{\Delta E} \rb^{2}\rbk  \frac{1}{e^{2\pi\Delta E/|a|} + 1}.
\eqe

We see that the above power radiated is characterized by a fermionic distribution as well as the half-integer indexed polynomial of multiplicity. The root cause of this change in statistics is the fact that we had poles of odd integer order rather than even integer order as in the transition rate. The first few normalized power functions, $\tilde{\mathcal{S}} = \mathcal{S}/G_{n}^{2}$, are then given by

\bqa
\tilde{\mathcal{S}}_{1}(\Delta E, a) &=& \frac{\Delta E^{2}}{2\pi}   \frac{ 1 + \frac{1}{4}  \lb \frac{a}{\Delta E} \rb^{2}}{e^{2\pi\Delta E/|a|} + 1} \non \\
\tilde{\mathcal{S}}_{2}(\Delta E, a) &=& \frac{\Delta E^{4}}{96\pi^{3}}   \frac{ 1 + \frac{5}{2}  \lb \frac{a}{\Delta E} \rb^{2} + \frac{9}{16}  \lb \frac{a}{\Delta E} \rb^{4}}{e^{2\pi\Delta E/|a|} + 1} \non \\
\tilde{\mathcal{S}}_{3}(\Delta E, a) &=& \frac{\Delta E^{6}}{11520\pi^{5}}   \frac{ 1 + \frac{35}{4}  \lb \frac{a}{\Delta E} \rb^{2} + \frac{259}{16}  \lb \frac{a}{\Delta E} \rb^{4}+ \frac{225}{64}  \lb \frac{a}{\Delta E} \rb^{6}}{e^{2\pi\Delta E/|a|} + 1} \non \\
\tilde{\mathcal{S}}_{4}(\Delta E, a) &=& \frac{\Delta E^{8}}{2580480\pi^{7}}   \frac{ 1 + 21  \lb \frac{a}{\Delta E} \rb^{2} + \frac{987}{8}  \lb \frac{a}{\Delta E} \rb^{4}+ \frac{3229}{16}  \lb \frac{a}{\Delta E} \rb^{6} + \frac{11025}{256}  \lb \frac{a}{\Delta E} \rb^{8}}{e^{2\pi\Delta E/|a|} + 1}.
\eqa

We note that the $n=1$ case, along with $\Delta E = 0$ as determined in [18], reproduces the known $a^{2}$ dependence for the power emitted by bremsstrahlung. The plots of the power radiated for both transitions up and down in Rindler energy can be found in Figs. 4 and 5 respectively. Note that the power radiated diverges in the case of a positive energy transitions. This reflects that inertially stable particles do not transition up in energy and radiate energy away. 

\begin{figure}[H]
\centering  
\includegraphics[,scale=.6]{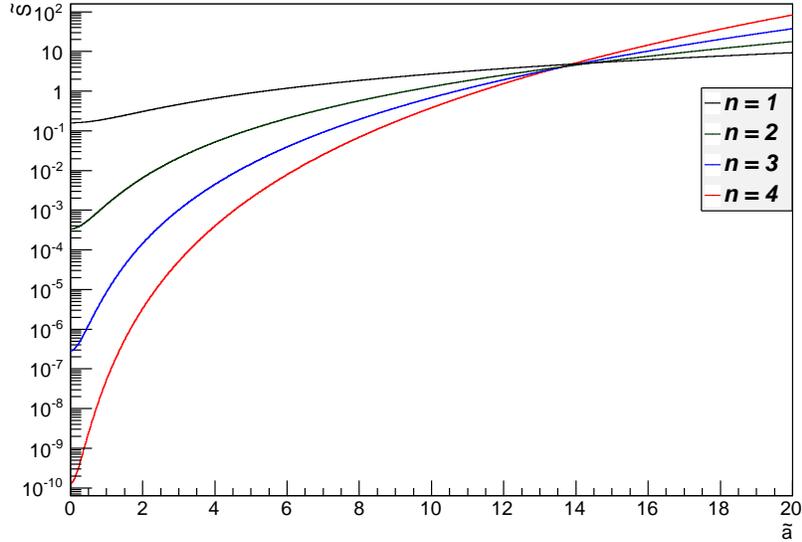}
\caption{The normalized power radiated, Eq. (67), with $\tilde{a} = a/\Delta E$ and $\Delta E=-1$.}
\end{figure} 

\begin{figure}[H]
\centering  
\includegraphics[,scale=.6]{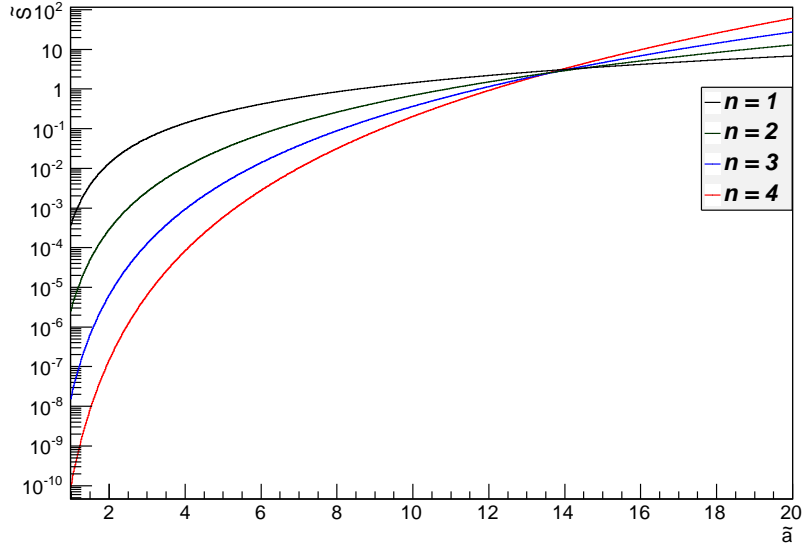}
\caption{The normalized power radiated, Eq. (67), with $\tilde{a} = a/\Delta E$ and $\Delta E=1$.}
\end{figure}

\section{Energy Spectra}

Utilizing the same prescription, we now endeavor to compute the energy spectrum of the $i$th Minkowski particle emitted in the transition process. The transition rate from Eq. (16) characterizes the probability per unit time that a transition will occur. The differential transition per unit energy then characterizes how the probability of emission changes with energy, i.e. the spectra. Therefore we begin by taking the derivative with respect to the $i$th particle's proper energy. Hence

\bqa
\frac{d\Gamma}{d\tilde{\omega_{i}}} &=& \frac{d}{d\tilde{\omega_{i}}} G_{n}^{2}\frac{1}{a} \int d\xi e^{-i\Delta E\xi /a} \prod_{m = 1}^{n}G_{m}^{\pm}[x',x]_{\eta} \non \\
 &=&  G_{n}^{2}\frac{1}{a} \int d\xi e^{-i\Delta E\xi /a}\frac{d}{d\tilde{\omega_{i}}}G_{i}^{\pm}[x',x]_{\eta} \prod_{m \neq i}^{n}G_{m}^{\pm}[x',x]_{\eta}.
\eqa

Again, we have separated out the relevant variant of the Wightman function for the observable we are calculating. Referring to Eqs. (34) and (38)  for the Wightman function and its derivative respectively, the above energy spectra reduce to a more convenient form. Hence

\bqa
\frac{d\Gamma}{d\tilde{\omega_{i}}} &=&  G_{n}^{2}\frac{1}{a} \int d\xi e^{-i\Delta E\xi /a}\frac{d}{d\tilde{\omega_{i}}}G_{i}^{\pm}[x',x]_{\eta} \prod_{m \neq i}^{n}G_{m}^{\pm}[x',x]_{\eta} \non \\
&=& G_{n}^{2}\frac{1}{a} \int d\xi e^{-i\Delta E\xi /a}\frac{\tilde{\omega}}{(2 \pi)^{2}}e^{-i\tilde{\omega}a\xi} \lbk -\frac{a^{2}}{(4 \pi)^{2}} \frac{1}{\sinh^{2}{[\xi/2-\sgn{(a)}i\epsilon]}} \rbk^{n-1} \non \\
&=& G_{n}^{2} \frac{\tilde{\omega}}{(2 \pi)^{2}} \lb \frac{ia}{4 \pi} \rb^{2(n-1)} \frac{1}{|a|} \int d\xi \frac{e^{-i(\Delta E + \tilde{\omega})\xi/|a|}}{ \sinh^{2(n-1)}{[\xi/2-i\epsilon]}}. 
\eqa

Note in the last line that we enforced the invariance of the above integral under a change in sign of the acceleration. We have encountered similar integrals for both the transition rate and the power radiated. It is worth noting that the difference is that the frequency variable which we are Fourier transforming with respect to has shifted via $\Delta E \rightarrow \Delta E + \tilde{\omega}$. We can evaluate the $n = 1$ case  quite easily at this point. Hence

\bqa
\lb \frac{d\Gamma}{d\tilde{\omega}}\rb_{n = 1} &=& G_{1}^{2} \frac{\tilde{\omega}}{(2 \pi)^{2}}   \frac{1}{|a|} \int d\xi   e^{-i(\Delta E+\tilde{\omega})\xi /|a|} \non \\
&=& G_{1}^{2} \frac{\tilde{\omega}}{2 \pi} \delta(\Delta E+\tilde{\omega}).
\eqa

The above expression is merely a statement analogous to Fermi's golden rule. Note that the presence of the delta function serves to enforce conservation of energy in the case of one particle emission. This implies that when there is only one Minkowski particle emitted that the radiated particle, as measured in an inertial frame instantaneously at rest with the accelerated particle, carries away the total change in the Rindler space energy. For the higher multiplicity cases we rid ourselves of the regulator and note the similar pole structure of order $2(n-1)$ when $\xi = 2\pi i \sigma$ with integer $\sigma \geq 1$. Finally, we make the same change of variables $w=e^{i\xi}$ to obtain

\bqa
\frac{d\Gamma}{d\tilde{\omega}} &=& G_{n}^{2} \frac{\tilde{\omega}}{(2 \pi)^{2}}  \lb \frac{ia}{4 \pi} \rb^{2(n-1)} \frac{1}{|a|} \int d\xi   \frac{e^{-i(\Delta E+\tilde{\omega})\xi /|a|}}{\sinh^{2(n-1)}{[\xi/2-i\epsilon]}} \non \\
&=& G_{n}^{2} \frac{\tilde{\omega}}{(2 \pi)^{2}}  \lb \frac{ia}{2 \pi} \rb^{2(n-1)} \frac{1}{|a|} \int d\xi   \frac{e^{-i(\Delta E+\tilde{\omega})\xi /|a|}}{\lbk e^{\xi/2 }- e^{-\xi/2}\rbk^{2(n-1)}} \non \\
&=& G_{n}^{2} \frac{\tilde{\omega}}{(2 \pi)^{2}}  \lb \frac{ia}{2 \pi} \rb^{2(n-1)} \frac{1}{|a|} \int dw   \frac{w^{-i(\Delta E+\tilde{\omega})/|a|+ n - 2} }{\lbk w- 1\rbk^{2(n-1)}}.
\eqa 

The above integral can be evaluated using the integration formula from Eq. (44) provided we make the relevant identification of the indices $\beta$, $\gamma$, and $\delta$. Moreover, since the $\gamma$ is an integer and $\delta$ is an even integer, the Pochhammer identity holds as well. Thus, by making the identification of $n \rightarrow n-1$ from the transition rate, we may merely quote the final form of the spectra. Hence

\bqa
\frac{d\Gamma}{d\tilde{\omega}} &=&  \frac{G^{2}_{n}\tilde{\omega}}{(2 \pi)^{2}} \lb \frac{\Delta E+\tilde{\omega}}{\pi}\rb^{2n-3} \frac{1}{(4n-6)!!}  \prod_{k = 0}^{n-2}\lbk 1  + k^{2} \lb \frac{ a}{\Delta E+\tilde{\omega}} \rb^{2} \rbk  \frac{1}{e^{2\pi(\Delta E+\tilde{\omega})/|a|} - 1}.
\eqa

Here we find a bosonic thermal distribution with a chemical potential and a polynomial of multiplicity characterized by an integer index. It is interesting to note that the total change in Rindler space energy is identified as the chemical potential of the thermal bath. Recalling the above spectra gives the probability of emission per unit energy per unit time, we note that it needs to be normalized via $\frac{1}{\Gamma}\frac{d\Gamma}{d \tilde{\omega}} =  \mathcal{N}$. This serves to scale the overall spectra and remove the effective differential time averaging. Below we write out the first few spectra normalized to the coupling via $\frac{1}{G_{n}^{2}}\frac{d\Gamma}{d \tilde{\omega}}  = \tilde{\mathcal{N}}$ as in the previous sections. Hence

\bqa
\tilde{\mathcal{N}}_{1}(\Delta E, a, \tilde{\omega}) &=& \frac{\tilde{\omega}}{2 \pi} \delta(\Delta E+\tilde{\omega}) \non \\
\tilde{\mathcal{N}}_{2}(\Delta E, a, \tilde{\omega}) &=& \frac{\tilde{\omega} (\Delta E+\tilde{\omega})}{8 \pi^{3}}\frac{1}{e^{2\pi(\Delta E+\tilde{\omega})/|a|} - 1} \non \\
\tilde{\mathcal{N}}_{3}(\Delta E, a, \tilde{\omega}) &=& \frac{\tilde{\omega} (\Delta E+\tilde{\omega})^{3}}{192 \pi^{5}}\frac{1+\lb \frac{a}{\Delta E+\tilde{\omega}}  \rb^{2}}{e^{2\pi(\Delta E+\tilde{\omega})/|a|} - 1} \non \\
\tilde{\mathcal{N}}_{4}(\Delta E, a, \tilde{\omega}) &=& \frac{\tilde{\omega} (\Delta E+\tilde{\omega})^{5}}{15360 \pi^{7}}\frac{1+ 5 \lb \frac{a}{\Delta E+\tilde{\omega}}  \rb^{2} + 4 \lb \frac{a}{\Delta E+\tilde{\omega}}  \rb^{4}}{e^{2\pi(\Delta E+\tilde{\omega})/|a|} - 1} \non \\
\tilde{\mathcal{N}}_{5}(\Delta E, a, \tilde{\omega}) &=& \frac{\tilde{\omega} (\Delta E+\tilde{\omega})^{7}}{2580480 \pi^{9}}\frac{1+ 14 \lb \frac{a}{\Delta E+\tilde{\omega}}  \rb^{2} + 49 \lb \frac{a}{\Delta E+\tilde{\omega}}  \rb^{4} + 36 \lb \frac{a}{\Delta E+\tilde{\omega}}  \rb^{6}}{e^{2\pi(\Delta E+\tilde{\omega})/|a|} - 1}.
\eqa 

In Figs. 6 and 7 we show the spectra normalized by the transition rate, i.e. $\mathcal{N}_{i}$, of the $i$th particle emitted for both transitions down and up in Rindler energy respectively. It is interesting to note that regardless of the sign of the transition energy, we find Planck-like spectra for all multiplicities greater than 1. 

\begin{figure}[H]
\centering  
\includegraphics[,scale=.6]{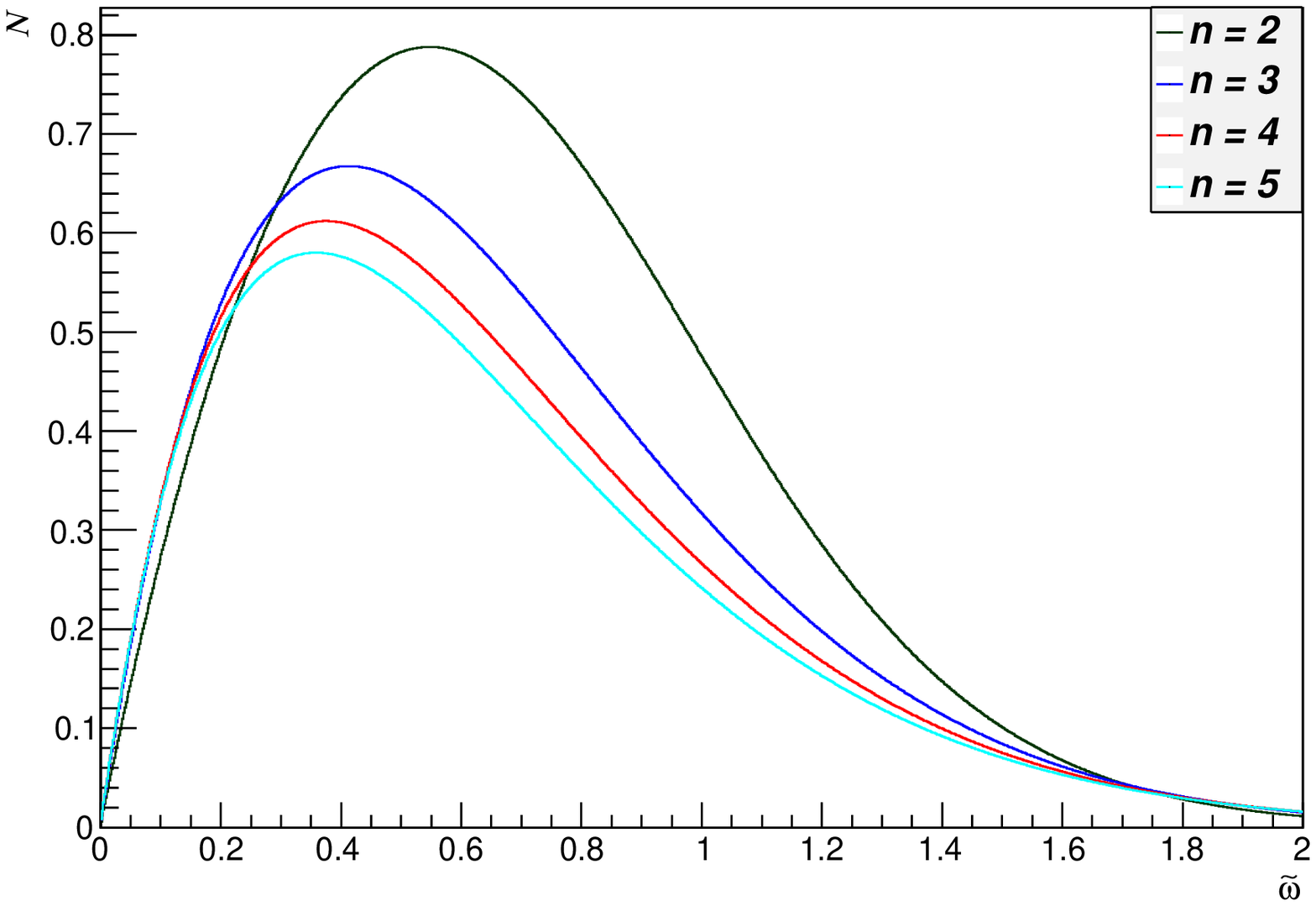}
\caption{The normalized spectra, $\mathcal{N}_{i} = \frac{1}{\Gamma}\frac{d\Gamma}{d \tilde{\omega_{i}}}$ , with $a = 1$ and $\Delta E=-1$.}
\end{figure} 

\begin{figure}[H]
\centering  
\includegraphics[,scale=.6]{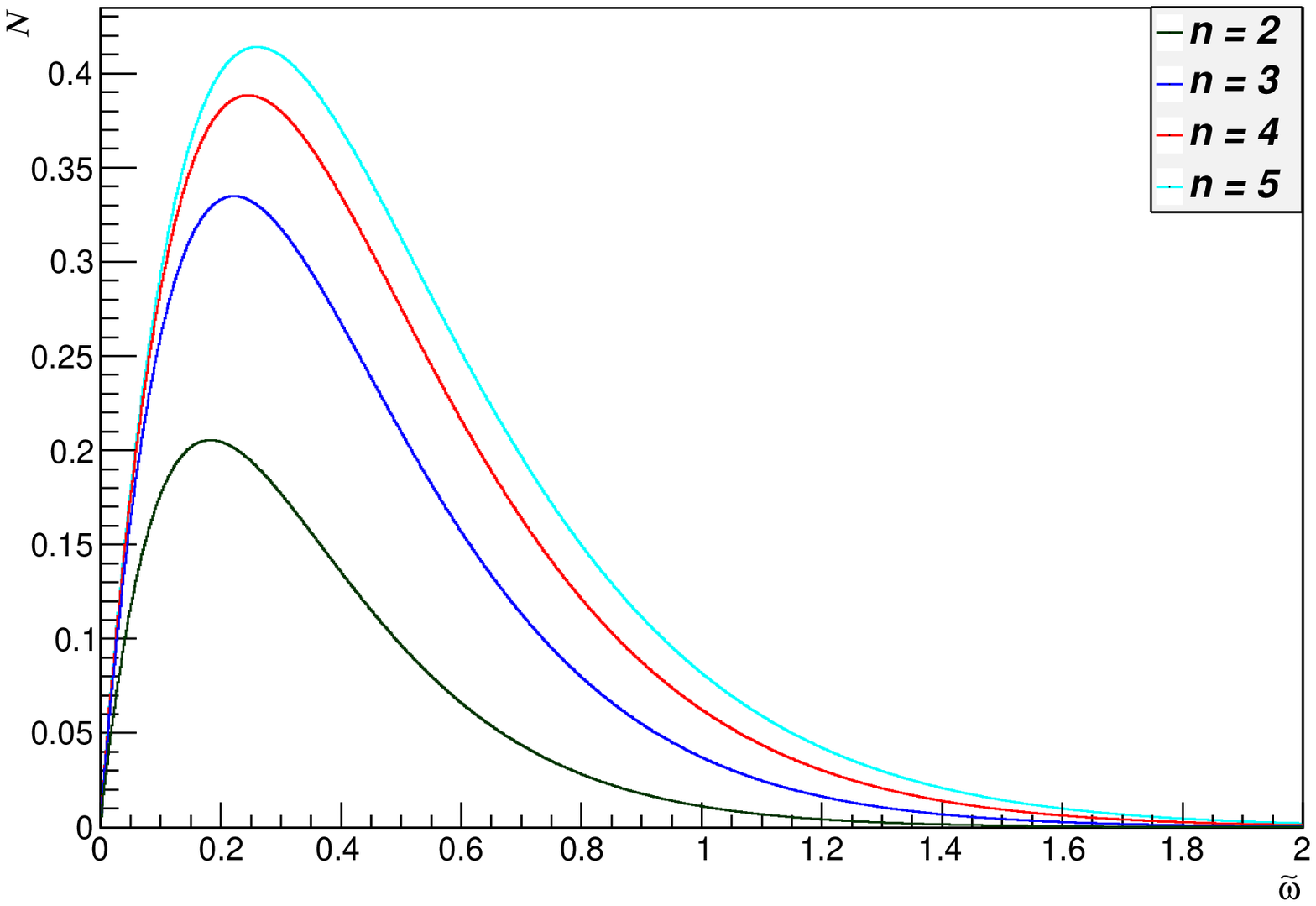}
\caption{The normalized spectra, $\mathcal{N}_{i} = \frac{1}{\Gamma}\frac{d\Gamma}{d \tilde{\omega_{i}}}$, with $a = 1$ and $\Delta E=1$.}
\end{figure} 

\section{Displacement Law}

To better characterize the spectra, we can also look at the peak energy of the emitted Minkowski particle for each multiplicity greater than 1. In the interest of determining the maximum of each peak via setting the derivative equal to zero, we can drop all prefactors and focus only on the energy dependence. Labeling the spectra polynomial of multiplicity $\mathcal{M}_{n} =  \prod_{k = 0}^{n-2}\lbk 1  + k^{2} \lb \frac{ a}{\Delta E+\tilde{\omega}} \rb^{2} \rbk$ we have

\bqa
\tilde{\mathcal{N}} &\sim &   \frac{\tilde{\omega}(\Delta E+\tilde{\omega})^{2n-3}\mathcal{M}_{n}}{e^{2\pi(\Delta E+\tilde{\omega})/|\tilde{a}_{\eta}|}-1}.
\eqa

Then, by taking the derivative with respect to $\tilde{\omega}$ and setting it equal to zero, we find

\bqa
\lbk (\Delta E+\tilde{\omega}) + \tilde{\omega}(2n-3)+ \tilde{\omega}(\Delta E+\tilde{\omega})\frac{\mathcal{M'}_{n}}{\mathcal{M}_{n}}  \rbk \lb e^{2\pi(\Delta E+\tilde{\omega})/|\tilde{a}_{\eta}|}-1 \rb \non \\
- \tilde{\omega}(\Delta E+\tilde{\omega})\frac{2 \pi}{|a|}e^{2\pi(\Delta E+\tilde{\omega})/|\tilde{a}_{\eta}|} &=& 0 \non \\
\frac{xe^{x}}{e^{x} - 1} - \lbk \frac{1}{1-\frac{2 \pi \Delta E}{|a|x}} +(2n - 3)+\frac{|a|x}{2 \pi} \frac{\mathcal{M'}_{n}}{\mathcal{M}_{n}} \rbk &=& 0.
\eqa

Note in the last line that we defined the dimensionless parameter $x = \frac{2 \pi}{|a|}(\Delta E + \tilde{\omega})$. Now we must evaluate the logarithmic derivative of the polynomial of multiplicity. Hence

\bqa
\frac{\mathcal{M'}_{n}}{\mathcal{M}_{n}} &=& \frac{d}{d\tilde{\omega}}\ln{\mathcal{M}_{n}} \non \\
&=& \frac{d}{d\tilde{\omega}}\ln{\prod_{k = 0}^{n-2}\lbk 1  + k^{2} \lb \frac{ a}{\Delta E+\tilde{\omega}} \rb^{2} \rbk} \non \\
&=& \frac{d}{d\tilde{\omega}}\sum_{k = 0}^{n-2}\ln{\lbk 1  + k^{2} \lb \frac{ a}{\Delta E+\tilde{\omega}} \rb^{2} \rbk} \non \\
&=& -\frac{2a^{2}}{(\Delta E + \tilde{\omega})^{3}}\sum_{k = 0}^{n-2}\frac{k^{2}}{1  + k^{2} \lb \frac{ a}{\Delta E+\tilde{\omega}} \rb^{2}}  \non \\
&=& -\frac{2}{|a|}\lb \frac{2 \pi}{x} \rb^{3}\sum_{k = 0}^{n-2}\frac{k^{2}}{1  + k^{2} \lb \frac{2 \pi}{x} \rb^{2}}.
\eqa

Then, combining the logarithmic derivative with the Eq. (75), we obtain our numerically solvable displacement law which allows us to determine the peak energy of the emitted Minkowski particles. Hence

\bqa
\frac{xe^{x}}{e^{x} - 1} - \lbk \frac{1}{1-\frac{2 \pi \Delta E}{|a|x}} +(2n - 3)-2\lb \frac{2 \pi}{x}\rb^{2} \sum_{k = 0}^{n-2}\frac{k^{2}}{1  + k^{2} \lb \frac{2 \pi}{x} \rb^{2}}  \rbk &=& 0.
\eqa

Then, in terms of the numerically solved $x$, we find that the peak energy is given by

\bqe
\tilde{\omega} = x \frac{|a|}{2 \pi} - \Delta E.
\eqe

Thus we have shown that the emitted particle's energy, in the limit of high acceleration or zero change in the Rindler space energy, is directly proportional to the accelerated temperature, i.e. $\tilde{\omega} = xt_{a}$. This is in agreement with Wien's displacement law but now with a more general transcendental equation to determine the displacement constant. Moreover, this establishes the fact that acceleration has an energy associated with it that is quantum mechanical in nature. By reinserting all relevant physical constants, we find the energy of acceleration $E_{a}$ is given by

\bqe
E_{a} = \frac{xa\hbar}{2 \pi c}.  
\eqe

The acceleration dependence of the energy of the emitted particles would have a clear signature at sufficiently high accelerations. Advanced experimental systems could be coming online in the coming years  that may be able to verify these effects [13].

\section{Conclusions}
In this paper we have established a computational framework capable of computing various observables of acceleration-induced particle physics processes. To better analyze physically realistic settings, a time-dependent formalism was developed to compute the spacetime quantities that go into the Wightman functions and its variants. These were then used to compute the transition rate, multiplicity, power radiated, energy spectra, and displacement law for accelerated decays and excitations of arbitrary final state multiplicity. We found the transition rate, power, and spectra are characterized by integer and half-integer indexed polynomials and thermal distributions of both bosonic and fermionic statistics. For the spectra, we found the total change in Rindler space energy plays the role of a chemical potential of the thermal bath. The displacement law for the spectra predicts the peak energy of the emitted Minkowski particles have a proper energy proportional to the accelerated temperature. This implies that there exists an energy, of quantum mechanical origin, associated with acceleration.  

\section*{Acknowledgments}
The author wishes to thank Luiz da Silva and Daniel Agterberg for many valuable discussions. This research was supported, in part, by the Leonard E. Parker Center for Gravitation, Cosmology, and Astrophysics and the University of Wisconsin at Milwaukee Department of Physics.

\goodbreak

\end{document}